\documentclass[superscriptaddress,showpacs,amsmath,amssymb,aps,prc,floatfix,twocolumn,showkeys]{revtex4-1}

\usepackage{graphicx}
\usepackage{dcolumn}
\usepackage{bm}
\usepackage[mathlines]{lineno}

\usepackage{slashbox}
\begin{document}

\title{Induced polarization of ${\Lambda} (1116)$ in kaon electroproduction}
\newcommand*{\ANL}{Argonne National Laboratory, Argonne, Illinois 60439}
\newcommand*{\ANLindex}{1}
\affiliation{\ANL}
\newcommand*{\ASU}{Arizona State University, Tempe, Arizona 85287-1504}
\newcommand*{\ASUindex}{2}
\affiliation{\ASU}
\newcommand*{\CSUDH}{California State University, Dominguez Hills, Carson, CA 90747}
\newcommand*{\CSUDHindex}{3}
\affiliation{\CSUDH}
\newcommand*{\CMU}{Carnegie Mellon University, Pittsburgh, Pennsylvania 15213}
\newcommand*{\CMUindex}{4}
\affiliation{\CMU}
\newcommand*{\CUA}{Catholic University of America, Washington, D.C. 20064}
\newcommand*{\CUAindex}{5}
\affiliation{\CUA}
\newcommand*{\SACLAY}{CEA, Centre de Saclay, Irfu/Service de Physique Nucl\'eaire, 91191 Gif-sur-Yvette, France}
\newcommand*{\SACLAYindex}{6}
\affiliation{\SACLAY}
\newcommand*{\CNU}{Christopher Newport University, Newport News, Virginia 23606}
\newcommand*{\CNUindex}{7}
\affiliation{\CNU}
\newcommand*{\UCONN}{University of Connecticut, Storrs, Connecticut 06269}
\newcommand*{\UCONNindex}{8}
\affiliation{\UCONN}
\newcommand*{\FU}{Fairfield University, Fairfield CT 06824}
\newcommand*{\FUindex}{9}
\affiliation{\FU}
\newcommand*{\FIU}{Florida International University, Miami, Florida 33199}
\newcommand*{\FIUindex}{10}
\affiliation{\FIU}
\newcommand*{\FSU}{Florida State University, Tallahassee, Florida 32306}
\newcommand*{\FSUindex}{11}
\affiliation{\FSU}
\newcommand*{\GWUI}{The George Washington University, Washington, DC 20052}
\newcommand*{\GWUIindex}{12}
\affiliation{\GWUI}
\newcommand*{\ISU}{Idaho State University, Pocatello, Idaho 83209}
\newcommand*{\ISUindex}{13}
\affiliation{\ISU}
\newcommand*{\INFNFE}{INFN, Sezione di Ferrara, 44100 Ferrara, Italy}
\newcommand*{\INFNFEindex}{14}
\affiliation{\INFNFE}
\newcommand*{\INFNFR}{INFN, Laboratori Nazionali di Frascati, 00044 Frascati, Italy}
\newcommand*{\INFNFRindex}{15}
\affiliation{\INFNFR}
\newcommand*{\INFNGE}{INFN, Sezione di Genova, 16146 Genova, Italy}
\newcommand*{\INFNGEindex}{16}
\affiliation{\INFNGE}
\newcommand*{\INFNRO}{INFN, Sezione di Roma Tor Vergata, 00133 Rome, Italy}
\newcommand*{\INFNROindex}{17}
\affiliation{\INFNRO}
\newcommand*{\ORSAY}{Institut de Physique Nucl\'eaire ORSAY, Orsay, France}
\newcommand*{\ORSAYindex}{18}
\affiliation{\ORSAY}
\newcommand*{\ITEP}{Institute of Theoretical and Experimental Physics, Moscow, 117259, Russia}
\newcommand*{\ITEPindex}{19}
\affiliation{\ITEP}
\newcommand*{\JMU}{James Madison University, Harrisonburg, Virginia 22807}
\newcommand*{\JMUindex}{20}
\affiliation{\JMU}
\newcommand*{\KNU}{Kyungpook National University, Daegu 702-701, Republic of Korea}
\newcommand*{\KNUindex}{21}
\affiliation{\KNU}
\newcommand*{\LPSC}{LPSC, Universite Joseph Fourier, CNRS/IN2P3, INPG, Grenoble, France}
\newcommand*{\LPSCindex}{22}
\affiliation{\LPSC}
\newcommand*{\UNH}{University of New Hampshire, Durham, New Hampshire 03824-3568}
\newcommand*{\UNHindex}{23}
\affiliation{\UNH}
\newcommand*{\NSU}{Norfolk State University, Norfolk, Virginia 23504}
\newcommand*{\NSUindex}{24}
\affiliation{\NSU}
\newcommand*{\OHIOU}{Ohio University, Athens, Ohio  45701}
\newcommand*{\OHIOUindex}{25}
\affiliation{\OHIOU}
\newcommand*{\ODU}{Old Dominion University, Norfolk, Virginia 23529}
\newcommand*{\ODUindex}{26}
\affiliation{\ODU}
\newcommand*{\RPI}{Rensselaer Polytechnic Institute, Troy, New York 12180-3590}
\newcommand*{\RPIindex}{27}
\affiliation{\RPI}
\newcommand*{\MSU}{Skobeltsyn Institute of Nuclear Physics, Lomonosov Moscow State University, 119234 Moscow, Russia}
\newcommand*{\MSUindex}{28}
\affiliation{\MSU}
\newcommand*{\SCAROLINA}{University of South Carolina, Columbia, South Carolina 29208}
\newcommand*{\SCAROLINAindex}{29}
\affiliation{\SCAROLINA}
\newcommand*{\JLAB}{Thomas Jefferson National Accelerator Facility, Newport News, Virginia 23606}
\newcommand*{\JLABindex}{30}
\affiliation{\JLAB}
\newcommand*{\UTFSM}{Universidad T\'{e}cnica Federico Santa Mar\'{i}a, Casilla 110-V Valpara\'{i}so, Chile}
\newcommand*{\UTFSMindex}{31}
\affiliation{\UTFSM}
\newcommand*{\EDINBURGH}{Edinburgh University, Edinburgh EH9 3JZ, United Kingdom}
\newcommand*{\EDINBURGHindex}{32}
\affiliation{\EDINBURGH}
\newcommand*{\GLASGOW}{University of Glasgow, Glasgow G12 8QQ, United Kingdom}
\newcommand*{\GLASGOWindex}{33}
\affiliation{\GLASGOW}
\newcommand*{\VT}{Virginia Polytechnic Institute and State University, Blacksburg, Virginia   24061-0435}
\newcommand*{\VTindex}{34}
\affiliation{\VT}
\newcommand*{\VIRGINIA}{University of Virginia, Charlottesville, Virginia 22901}
\newcommand*{\VIRGINIAindex}{35}
\affiliation{\VIRGINIA}
\newcommand*{\WM}{College of William and Mary, Williamsburg, Virginia 23187-8795}
\newcommand*{\WMindex}{36}
\affiliation{\WM}
\newcommand*{\YEREVAN}{Yerevan Physics Institute, 375036 Yerevan, Armenia}
\newcommand*{\YEREVANindex}{37}
\affiliation{\YEREVAN}
\newcommand*{\Rich}{University of Richmond, Richmond, Virginia 23221}
\newcommand*{\Richindex}{38}
\affiliation{\Rich}

\newcommand*{\NOWNONE}{No address available}
\newcommand*{\NOWORSAY}{Institut de Physique Nucl\'eaire ORSAY, Orsay, France}
\newcommand*{\NOWODU}{Old Dominion University, Norfolk, Virginia 23529}
\newcommand*{\NOWGLASGOW}{University of Glasgow, Glasgow G12 8QQ, United Kingdom}

\author{M. Gabrielyan}\affiliation{\FIU}
\author{B. A. Raue}\affiliation{\FIU}
\author{D. S. Carman}\affiliation{\JLAB}
\author {K.~Park}\altaffiliation[Current address: ]{\NOWODU}\affiliation{\JLAB}\affiliation{\KNU}
\author {K.P. ~Adhikari}\affiliation{\ODU}
\author {D.~Adikaram} \affiliation{\ODU}
\author {M.J.~Amaryan}\affiliation{\ODU}
\author {S. ~Anefalos~Pereira} \affiliation{\INFNFR}
\author {H.~Avakian} \affiliation{\JLAB}
\author {J.~Ball} \affiliation{\SACLAY}
% ...

\collaboration{The CLAS Collaboration}
\noaffiliation

\date{\today}

\begin{abstract}
  We have measured the induced polarization of the $\Lambda$(1116) in
  the reaction $ep \to e'K^+\Lambda$, detecting the scattered $e'$ and $K^+$ in the final state
  along with the proton from the decay $\Lambda \to p \pi^-$. The present study used the CEBAF
  Large Acceptance Spectrometer (CLAS), which allowed for a large
  kinematic acceptance in invariant energy $W$ ($1.6\leq W \leq 2.7$~GeV) and covered
  the full range of the kaon production angle at an average momentum transfer
  $Q^2$=1.90~GeV$^2$. In this experiment a 5.50~GeV electron beam
  was incident upon an unpolarized liquid-hydrogen target.  
  We have mapped out the $W$ and kaon production angle dependencies of the induced polarization 
  and found striking differences from photoproduction data over most of the kinematic range studied. 
  However, we also found that the induced polarization is essentially $Q^2$ independent in our kinematic
  domain, suggesting that 
  somewhere below the $Q^2$ covered here there must be a strong $Q^2$ dependence.  Along with
  previously published photo- and electroproduction cross sections and
  polarization observables, these data
  are needed for the development of models, such as effective field
  theories, and as input to coupled-channel analyses that can provide evidence of
  previously unobserved $s$-channel resonances.
\end{abstract}

\keywords{kaon, hyperon, electroproduction, polarization}
\pacs{13.40.-f,13.60Rj,13.88.+e,14.20.Gk,14.20Jn}

\maketitle

\section{Introduction}
\label{sec-intro}

The strange quark plays an important role in understanding the strong
interactions of the nucleon~\cite{Azn2013,Mart2010,Capstick2000}.  The investigation of strangeness production in
both photo- and electroproduction reactions has been carried out since the
1970s, but there is still no comprehensive model describing the reaction
mechanism. This is due, in part, to the difficulties encountered in modeling the
strong interaction in the energy range of excited baryon
masses.  As such, the problem has been approached through the use of effective
field theories~\cite{haber,MB-d13,saghai_aip,delaPuente,maxwell:12,maxwell:12-2}, Regge
models~\cite{Guidal97,Guidal2000,Guidal2003}, hybrid Regge-plus-resonance
(RPR) models \cite{corthals,tom-vrancx}, and more recently, through
coupled-channel analyses \cite{Anisovich2012,Penner,Doring,Kamano}. All of these
methods require large and precise data sets in order to constrain fitting
parameters. The work presented in this paper is part of a larger program being
carried out by the CLAS Collaboration at Jefferson Lab to determine cross
sections and polarization observables in kaon photo- and electroproduction over a broad kinematic
range, which can then be used as input to constrain the aforementioned models.

An important part of these efforts is the identification of nucleon
resonances that couple to the $K^+\Lambda$ final state. Constituent quark models \cite{capstick:1993} predict the existence
of excited nucleon states, many of which have yet to be observed
experimentally. Many of the data on nucleon resonances come from $\pi N \to N^* \to \pi N$
reactions. However, because the density of states for this channel is high, unambiguously identifying the
signal for a relatively weak or broad resonance is difficult.  To fully
understand the production and decay of excited baryon states, other reaction
channels must be explored, such as electromagnetic  production with decay via kaon emission. 

Strangeness production experiments using hadronic or electromagnetic beams on
various nuclear targets have been carried out since the 1970s, but only in the
past decade have high-precision data on a large number of observables became
available. Data on differential cross sections and spin observables for $KY$
photoproduction have been published by the
SAPHIR~\cite{Tran:1998qw,Goers,glander},
LEPS~\cite{Zegers,Sumihama:2005er,Kohri}, GRAAL
\cite{Lleres2007,dangelo,Lleres2009}, and
CLAS~\cite{McNabb:2003nf,Bradford:2005pt,Bradford:2006ba,mccracken,dey:2010}
Collaborations. Together, these data cover the full range of $\cos \theta_K^{CM}$
and invariant energy $W$ from 1.6 to 2.8~GeV. The recent photoproduction results from
CLAS~\cite{mccracken,dey:2010} not only extended the existing $W$ range by 500~MeV, they
significantly improved the precision of the cross section and induced hyperon
polarization data.  These experiments have been essential in providing evidence for new excited states in the mass range
around 1900 MeV that are now included in the particle data tables \cite{pdg}.

High statistics data for $KY$ electroproduction are relatively sparse as
compared to photoproduction. Recently, the CLAS Collaboration published data on differential cross
sections and separated structure functions for the $K^+\Lambda$ and
$K^+\Sigma^0$ final states~\cite{Ambrozewicz:2006zj,Nasseripour:2008,Carman:2012}. 
These data cover the full kaon center-of-mass angular range with momentum transfer $Q^2$ from 0.5 to 3.9~GeV$^2$ and
$W$ from threshold to 2.6~GeV.  Differential cross sections and the separation
of the longitudinal and transverse structure functions in the
$ep \to e'K^+\Lambda$ and $ep \to e'K^+\Sigma^0$ reactions were published by the 
Jefferson Lab Hall~C Collaboration~\cite{mohring,Niculescu}. These data cover a $Q^2$ range from
0.5 to 2~GeV$^2$ at $W$=1.84~GeV. In a recent publication from Hall~A at
Jefferson Lab~\cite{Marius}, the longitudinal, $\sigma_L$, and transverse,
$\sigma_T$, structure functions were separated by the Rosenbluth technique at fixed
$W$ and $t$. These results cover the kinematic range for $Q^2$ from 1.90 to
2.35~GeV$^2$ and $W$ from 1.80 to 2.14~GeV. Recent
CLAS~\cite{Carman2003,Carman2009} beam-recoil polarization transfer data for
the exclusive $\vec{e}p \to e'K^+\vec{Y}$ (where $Y$ is either a $\Lambda$ or 
$\Sigma^0$) reaction have a wide kinematic coverage spanning
$Q^2$ from 0.7 to 5.4~GeV$^2$ and $W$ from 1.6 to 2.6~GeV.

Polarization observables possess a strong discriminatory power that can be used
for distinguishing between different theoretical models and their variants, 
for which the differential cross sections alone have proven to be insufficient. 
In this paper we present results for the induced polarization of the $\Lambda$ from
the reaction $ep \to e'K^+\vec{\Lambda}$.  These results, when added to
the world's database, will help to constrain model parameters of strangeness
production and potentially aid in the identification of missing $N^*$
resonances.

The organization of this paper is as follows. Section~\ref{sec-formalism} gives the
relevant formalism for the polarization observables presented in this paper. Section~\ref{sec-expt} 
contains the details of the experimental setup and describes all analysis cuts, data binning, 
corrections, and fitting procedures. Section~\ref{sec:systematics} contains a discussion of the sources of
systematic uncertainty on the polarization observables. Section~\ref{sec:results}
contains our results and discussion. Finally, Section~\ref{sec:conclusions} presents our conclusions.

\section{Formalism}
\label{sec-formalism}

\begin{figure}[t!]
  \includegraphics[width=0.49\textwidth] {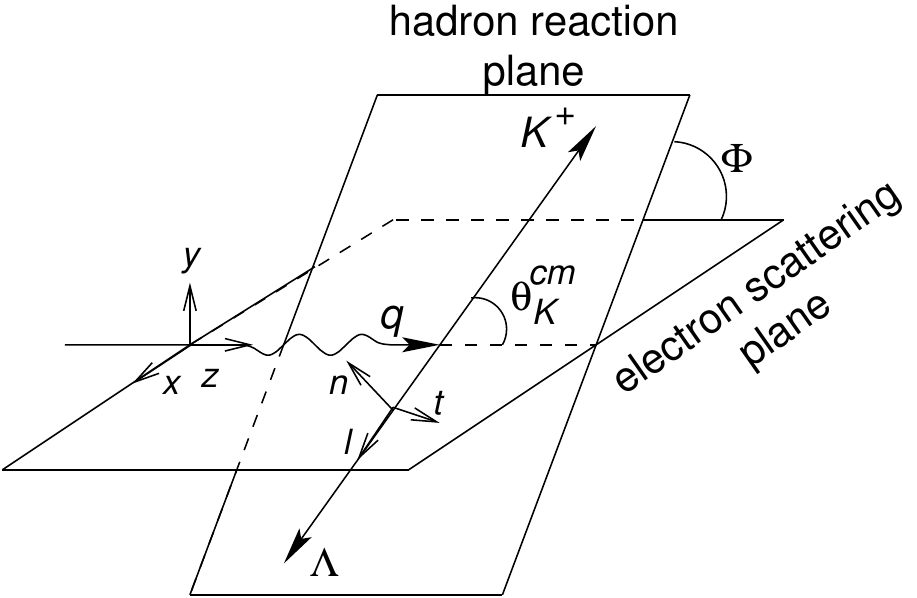}
  \caption[]{Kinematics for $K^+\Lambda$ electroproduction showing the angles and
    polarization axes in the CM reference frame.}
\label{fig-kin}
\end{figure}

For electroproduction, the reaction kinematics are uniquely defined by the set
of four variables ($Q^2$, $W$, $\cos\theta_K^{CM}$, $\Phi$), where $\theta_K^{CM}$ is
the kaon production angle in the virtual photon-proton center-of-mass (CM) frame defined in 
Fig.~\ref{fig-kin}, and $\Phi$ is the relative angle between the electron-scattering and the hadron-production
planes. $Q^2=-q^2$ is the squared four-momentum transfer of the virtual photon and $W=\sqrt{M_p^2+2M_p\nu-Q^2}$ is the invariant 
mass of the intermediate state, where $M_p$ is the proton mass and $\nu=E_e-E_{e'}$ is the difference between 
the incident ($E_e$) and scattered ($E_{e'}$) electron energies.

The cross sections and the polarization observables for pseudoscalar meson electroproduction can be expressed in terms of
36 non-zero response functions (see Table~\ref{tab-Rfn}) according to the framework of Ref.~\cite{Knochlein}. However, not all of these 
response functions are independent and a complete description of electroproduction requires only 11 independent measurements.
Some of these observables have already been measured as discussed in Section~\ref{sec-intro}.

The $K^+ \Lambda$ electroproduction cross section in the single-photon exchange approximation can
be expressed as a product of the virtual photon flux, $\Gamma$, and the virtual photoabsorption
cross section as
\begin{equation}
  \label{eq-csec1}
  \frac{d^5\sigma}{dE_{e'} d\Omega_{e'} d\Omega_K^{CM}} = \Gamma
  \frac{d^2\sigma_v}{d\Omega_K^{CM}}, 
\end{equation}
where  
\begin{equation}
  \label{gamma_flux}
  \Gamma=\frac{\alpha}{4\pi}\frac{W}{M_p^2E^2_e}(W^2-M_p^2) \left
    [\frac{1}{Q^2(1-\epsilon)} \right ].
\end{equation}
Here, $\alpha$ is the fine-structure constant and $\epsilon$ is the virtual photon polarization parameter given by
\begin{equation}
  \label{gpol}
  \epsilon = \left[ 1+2 \left( 1+\frac{\nu^2}{Q^2} \right) \tan^2 \frac{\theta_{e'}}{2} \right ]^{-1},
 \end{equation}
where $\theta_{e'}$ is the scattered electron laboratory polar angle.  The virtual photoabsorption cross section can be written in terms of the response functions $R_i^{\beta\alpha}$ as
\begin{widetext}
  \begin{eqnarray}
    \label{csec2}
    \frac{d^2\sigma_v}{d\Omega_K^{CM}} &=& K S_\alpha S_\beta \Bigl[
    R_T^{\beta\alpha}+\epsilon R_L^{\beta\alpha} +\sqrt{\epsilon (1+\epsilon)}(\
    ^cR_{LT}^{\beta\alpha}\cos \Phi +\ ^sR_{LT}^{\beta\alpha}\sin{\Phi})
    \nonumber \\ 
    &+&\epsilon(\ ^cR_{TT}^{\beta\alpha}\cos{2\Phi} +\
    ^sR_{TT}^{\beta\alpha}\sin{2\Phi}) \nonumber \\ 
    &+&h\sqrt{\epsilon (1-\epsilon)} (\ ^cR_{LT'}^{\beta\alpha}\cos{\Phi} +\
    ^sR_{LT'}^{\beta\alpha}\sin{\Phi}) +h\sqrt{1-\epsilon^2}R_{TT'}^{\beta\alpha} \Bigr]. 
  \end{eqnarray}
\end{widetext}
In this expression, the kinematic factor, $K=\frac{|\vec{p}_K|}{k_\gamma^{CM}}$, is the ratio of the 
kaon and virtual photon momenta in the CM frame and $h$ is the
electron-beam helicity. The superscripts $\alpha$ and $\beta$ refer to the
target and $\Lambda$ polarizations, respectively, where a sum over $\alpha$ and
$\beta$ is implied.  The $c$ and $s$ superscripts on the response functions
refer to the cosine or sine terms that accompany them. Only one of these is non-zero for a given
combination of $\alpha$ and $\beta$ as summarized in Table~\ref{tab-Rfn}.

The spin-projection operators are defined as:
\begin{eqnarray*} 
  S_\alpha &=& (1,{\mathbf S}),\\
  S_\beta &=& (1,{\mathbf S'}),
\end{eqnarray*}
with
\begin{eqnarray*} 
  {\mathbf S} &=& (\hat{S}_x,\hat{S}_y,\hat{S}_z),\\
  {\mathbf S}' &=& (\hat{S}_{x'},\hat{S}_{y'},\hat{S}_{z'}).
\end{eqnarray*}
The unprimed-coordinate system is associated with the
electron-scattering plane. It is defined with the $\hat{z}$ axis along the
virtual photon momentum vector $\vec{q}$, with $\hat{y}$ normal to the
electron-scattering plane, and $\hat{x}=\hat{y}\times\hat{z}$.  The
primed-coordinate system is associated with the hadron-plane
coordinates and is defined so that $\hat{z}'$ is along the kaon momentum vector
$\vec{p}_K$, with $\hat{y}'$ normal to the hadron production plane, and
$\hat{x}'=\hat{y}'\times\hat{z}'$.

In the simplest case, with nothing polarized, the contributions from the beam,
target and recoil polarizations vanish, and Eq.~\ref{csec2} reduces to
\begin{widetext}
  \begin{equation}
    \label{csec-nopol}
    \sigma_0 \equiv \left( \frac{d^2\sigma_v}{d\Omega_K^{CM}} \right)^{00}\!\! 
    = K \left[R_T^{00} + \epsilon R_L^{00} + \sqrt{\epsilon
        (1+\epsilon)}R_{LT}^{00} \cos \Phi + \epsilon R_{TT}^{00} \cos{2\Phi}
    \right], 
  \end{equation}
\end{widetext}
so that $KR_i^{00}=\sigma_i$ are the unpolarized cross-section components.

\begin{table}[b!]
\begin{ruledtabular}
\begin{tabular} {|c|c||c|c|c|c|c|c||c|c|c|} 
$\beta$ & $\alpha$ & $T$ & $L$ & $^cLT$ & $^sLT$ & $^cTT$ & $^sTT$ & $^cLT'$ & $^sLT'$ & $TT'$ \\ \hline \hline
-    & -   & $R_T^{00}$ & $R_L^{00}$ & $R_{LT}^{00}$ & 0 & $R_{TT}^{00}$ & 0 & 0 & $R_{LT'}^{00}$ & 0  \\ 
-    & $x$ & 0 & 0 & 0 & $R_{LT}^{0x}$ & 0 & $R_{TT}^{0x}$ & $R_{LT'}^{0x}$ & 0 & $R_{TT'}^{0x} $ \\  
-    & $y$ & $R_T^{0y}$ & $R_L^{0y}$ & $R_{LT}^{0y}$ & 0 & $\ddag$ & 0 & 0 & $R_{LT'}^{0y}$ & 0 \\  
-    & $z$ & 0 & 0 & 0 & $R_{LT}^{0z}$ & 0 & $R_{TT}^{0z}$ & $R_{LT'}^{0z}$ & 0 & $R_{TT'}^{0z} $ \\
$x'$ & -   & 0 & 0 & 0 & $R_{LT}^{x'0}$ & 0 & $R_{TT}^{x'0}$ & $R_{LT'}^{x'0}$ & 0 & $R_{TT'}^{x'0}$  \\ 
$y'$ & -   & $R_T^{y'0}$ & $\ddag$ & $\ddag$ & 0 & $\ddag$ & 0 & 0 & $\ddag$ & 0 \\ 
$z'$ & -   & 0 & 0 & 0 & $R_{LT}^{z'0}$ & 0 & $R_{TT}^{z'0}$ & $R_{LT'}^{z'0}$ & 0 & $R_{TT'}^{z'0}$ \\ 
$x'$ & $x$ & $R_T^{x'x}$ & $R_L^{x'x}$ & $R_{LT}^{x'x}$ & 0 & $\ddag$ &  0 & 0 & $R_{LT'}^{x'x}$ & 0 \\
$x'$ & $y$ & 0 & 0 & 0 & $\ddag$ & 0 & $\ddag$ & $\ddag$ & 0 & $\ddag$ \\
$x'$ & $z$ & $R_T^{x'z}$ & $R_L^{x'z}$ & $\ddag$ & 0 & $\ddag$ & 0 & 0 & $\ddag$  & 0 \\
$y'$ & $x$ & 0 & 0 & 0 & $\ddag$ & 0 & $\ddag$ & $\ddag$ & 0 & $\ddag$ \\ 
$y'$ & $y$ & $\ddag$ & $\ddag$ & $\ddag$ & 0 & $\ddag$ & 0 & 0 & $\ddag$ & 0 \\ 
$y'$ & $z$ & 0 & 0 & 0 & $\ddag$ & 0 & $\ddag$ & $\ddag$ & 0 & $\ddag$ \\ 
$z'$ & $x$ & $R_T^{z'x}$ & $\ddag$ & $R_{LT}^{z'x}$ & 0 & $\ddag$ & 0 & 0 & $R_{LT'}^{z'x}$ & 0 \\  
$z'$ & $y$ & 0 & 0 & 0 & $\ddag$ & 0 & $\ddag$ & $\ddag$ & 0 & $\ddag$ \\ 
$z'$ & $z$ & $R_T^{z'z}$ & $\ddag$ & $\ddag$ & 0 & $\ddag$ & 0 & 0 & $\ddag$ & 0 \\ 
\end{tabular}
\caption[]{Response functions for pseudoscalar meson electroproduction~\cite{Knochlein}. The target (recoil) 
polarization is related to the coordinate axes denoted by $\alpha$ ($\beta$) (see Fig.~\ref{fig-kin}). The 
last three columns are for a polarized electron beam. The symbol $\ddag$ indicates a response function 
that does not vanish but is related to other response functions.}
\label{tab-Rfn}
\end{ruledtabular}
\end{table}

During this experiment, a longitudinally polarized electron beam was incident upon an
unpolarized proton target, producing a polarized recoil hyperon. Summed over both helicities of the
incident electron beam Eq.~\ref{csec2} becomes
\begin{equation}
  \label{csec3}
  \frac{d^2\sigma_v}{d\Omega_K^{CM}} = \sigma_0 (1 + P^0_{x'}\hat{S}_{x'}
  + P^0_{y'}\hat{S}_{y'} + P^0_{z'}\hat{S}_{z'}),
\end{equation}
where the $P^0_{j'}$ terms (with $j' = x', y', z'$) are the induced hyperon polarization components
with respect to the primed coordinate system. These components can be expressed in terms of the
response functions as: 
\begin{eqnarray}
  \label{eqn-Pxp0}
  P_{x'}^0 &=& \frac{K}{\sigma_0}\left( \sqrt{\epsilon (1+\epsilon)}\ R_{LT}^{x' 0}\sin{\Phi} 
     +\epsilon\ R_{TT}^{x' 0}\sin{2\Phi}\right) \\ 
  P_{y'}^0 &=& \frac{K}{\sigma_0}\Bigl( R_T^{y' 0} + \epsilon R_L^{y' 0}
     + \sqrt{\epsilon (1+\epsilon)}\ R_{LT}^{y' 0}\cos{\Phi}  \nonumber \\ 
   &+& \epsilon\ R_{TT}^{y' 0}\cos{2\Phi}\Bigr) \nonumber \\
  P_{z'}^0\! &=&\! \frac{K}{\sigma_0}\left( \sqrt{\epsilon (1+\epsilon)}\
     R_{LT}^{z' 0}\sin{\Phi} +\epsilon\ R_{TT}^{z' 0}\sin{2\Phi}\right). \nonumber 
\end{eqnarray}
The integration of Eq.~\ref{csec3} over the full $\Phi$ range, which is necessary in this
experiment to reduce statistical uncertainties and to allow for finer binning in
$W$ and $\cos\theta_K^{CM}$, leads to 
\begin{eqnarray}
  \label{Icsec}
  \int^{2\pi}_0 \frac{d^2\sigma_v}{d\Omega_K^{CM}} d\Phi = 2\pi K(R_T^{00}&+&\epsilon R_L^{00}) 
  \nonumber \\
  (1+{\cal P}^0_{x'} S_{x'}&+&{\cal P}^0_{y'} S_{y'}+{\cal P}^0_{z'}S_{z'}),
\end{eqnarray}
where ${\cal P}^0_{j'}$ are $\Phi$-integrated polarization components in the primed coordinate
system,
\begin{eqnarray}
  \label{eqn-Pxp0_phi}
  {\cal P}_{x'}^0\! &=&\! 0,  \\
  {\cal P}_{y'}^0\! &=&\! \frac{K}{\sigma_0} \left( R_T^{y'0} + \epsilon R_L^{y'0}\right), \ {\rm and} \nonumber \\  
  {\cal P}_{z'}^0\! &=&\! 0. \nonumber
\end{eqnarray}
Eqs.~\ref{eqn-Pxp0_phi} show that only the normal component of the induced polarization survives the 
$\Phi$ integration. 

The coordinate system, (${\hat{t},\hat{n},\hat{l}}$), which was used in this analysis, is defined with 
$\hat{l}$ along the $\Lambda$ momentum direction ($\hat{l}=-\hat{z'}$), $\hat{n}$ normal to the hadron 
plane ($\hat{n} = \hat{y'}$), and $\hat{t}=-\hat{x'}$.  The polarization components in this system are 
given by
\begin{equation}
  {\cal P}^0_{t}=-{\cal P}^0_{x'} \hskip 0.25in {\cal P}^0_{n}={\cal P}^0_{y'} \hskip
  0.25in {\cal P}^0_{l}=-{\cal P}^0_{z'}. 
\end{equation}

\section{Experiment Description and Data Analysis}
\label{sec-expt}
\subsection{CLAS Spectrometer}
 
The CEBAF Large Acceptance Spectrometer (CLAS) \cite{Mecking:2003} was designed to
allow operation with both electron and photon beams, while providing an acceptance
for charged particles of approximately 50\% of $4\pi$ in solid angle. The large acceptance of
CLAS is crucial for investigations of multi-particle final states that
result from the decay of baryons and mesons.

\begin{figure}[hbtp]
  \includegraphics[width=0.49\textwidth]{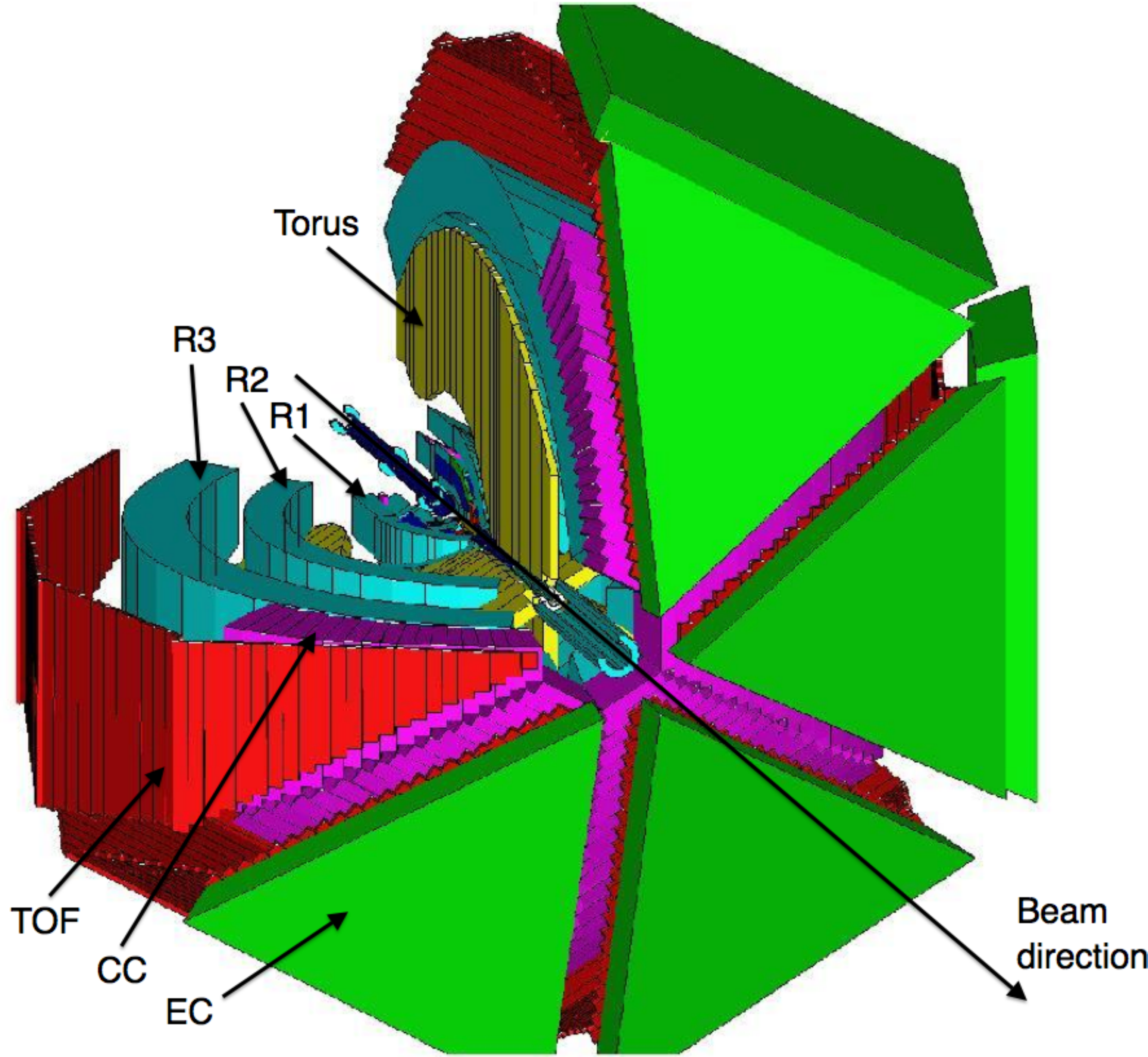}
  \caption{(Color online) Three-dimensional cut-away view of CLAS showing the drift chambers
    (R1, R2, and R3),  Cherenkov Counters (CC),  Time-of-Flight system
    (TOF), and Electromagnetic Calorimeters (EC). In this view, the beam
    enters the picture from the upper left corner and travels down the center of the detector. The 
    detector is roughly 10~m in diameter.}
  \label{clas_schem}
\end{figure}

CLAS was divided into six identical sectors by superconducting coils that
produced an approximately toroidal magnetic field about the beam axis. The field
was set at 60\% of its maximum for this experiment.  Each of the six CLAS
sectors was equipped with an identical set of detectors: three layers of drift
chambers (DC) \cite{dc_pp} for charged particle tracking and momentum
reconstruction, Cherenkov counters (CC)~\cite{cc_pp} for electron identification
and triggering, scintillation counters for time of flight (TOF)~\cite{tof_pp}
measurements and charged particle identification, and electromagnetic
calorimeters (EC)~\cite{ec_cal_pp} for electron identification and triggering.
The CLAS kinematic acceptance for this experiment was $0.8 \leq
Q^2\leq 3.5$~GeV$^2$ and $1.6\leq W \leq 2.7$~GeV, with hadron scattering
angles from 8$^\circ$ to 142$^\circ$ and electron scattering angles from
8$^\circ$ to 45$^\circ$. Fig.~\ref{clas_schem} illustrates a schematic view of
the CLAS detector subsystems.

In this experiment, a 5.50~GeV longitudinally polarized electron beam with an average beam
current of 7~nA was incident upon an unpolarized liquid-hydrogen target. The
target was 5.0~cm long and positioned 25~cm upstream of the nominal CLAS
center. The average luminosity was about $1 \times 10^{34}$~cm$^{-2}$s$^{-1}$. Event
readout was triggered by a coincidence between a CC hit and an EC hit in a
single sector, generating an event rate of about 2~kHz. The live-time corrected integrated 
luminosity of this data set is 11~fb$^{-1}$, and the data set for this analysis contained 
$\sim 1 \times 10^5$ $e'K^+\Lambda$ events.

\subsection{Event Identification}
\label{eid}

The trigger configuration ensured that all events had an electron
candidate. Electron candidates were also required to have a valid track in the DC
corresponding to a negatively charged particle and a hit in the TOF system that
coincided in time with the hit in the EC. The events for which these conditions
were not satisfied, were rejected in the offline analysis during event
reconstruction.  Additional cuts applied to improve the electron identification
included geometrical fiducial cuts, which made sure that electrons hit a region of CLAS
with a relatively flat acceptance, target-vertex cuts, which ensured that the
scattered electron came from the target, and EC fiducial cuts, which ensured complete
energy deposition in the calorimeter.

This analysis required the detection of a kaon and a proton from the $\Lambda$
decay along with the electron.  Hadrons were required to have a valid track in
the DC corresponding to a positively charged particle and a hit in the TOF
system that coincided in time. Hadrons were identified using the
time-of-flight difference $\Delta t = t_1-t_2$, where $t_1$ is the measured
time of flight from the interaction vertex position to the hit TOF paddle and
$t_2$ is the time for a particle with an assumed mass to travel the same
distance.  The time $t_2$ was calculated as
\begin{equation}
  \label{t2}
  t_2 = \frac{d}{c\beta_2},
\end{equation}
where $d$ is the measured flight path length and  
\begin{equation}
  \label{beta2}
  {\beta_2} = {\frac{p} {\sqrt{(m_2 c)^2+p^2}}}.
\end{equation}
Here, $m_2$ is the assumed particle mass and $p$ is the measured particle
momentum.  
\begin{figure*}[hbtp]
  \includegraphics[height=7cm]{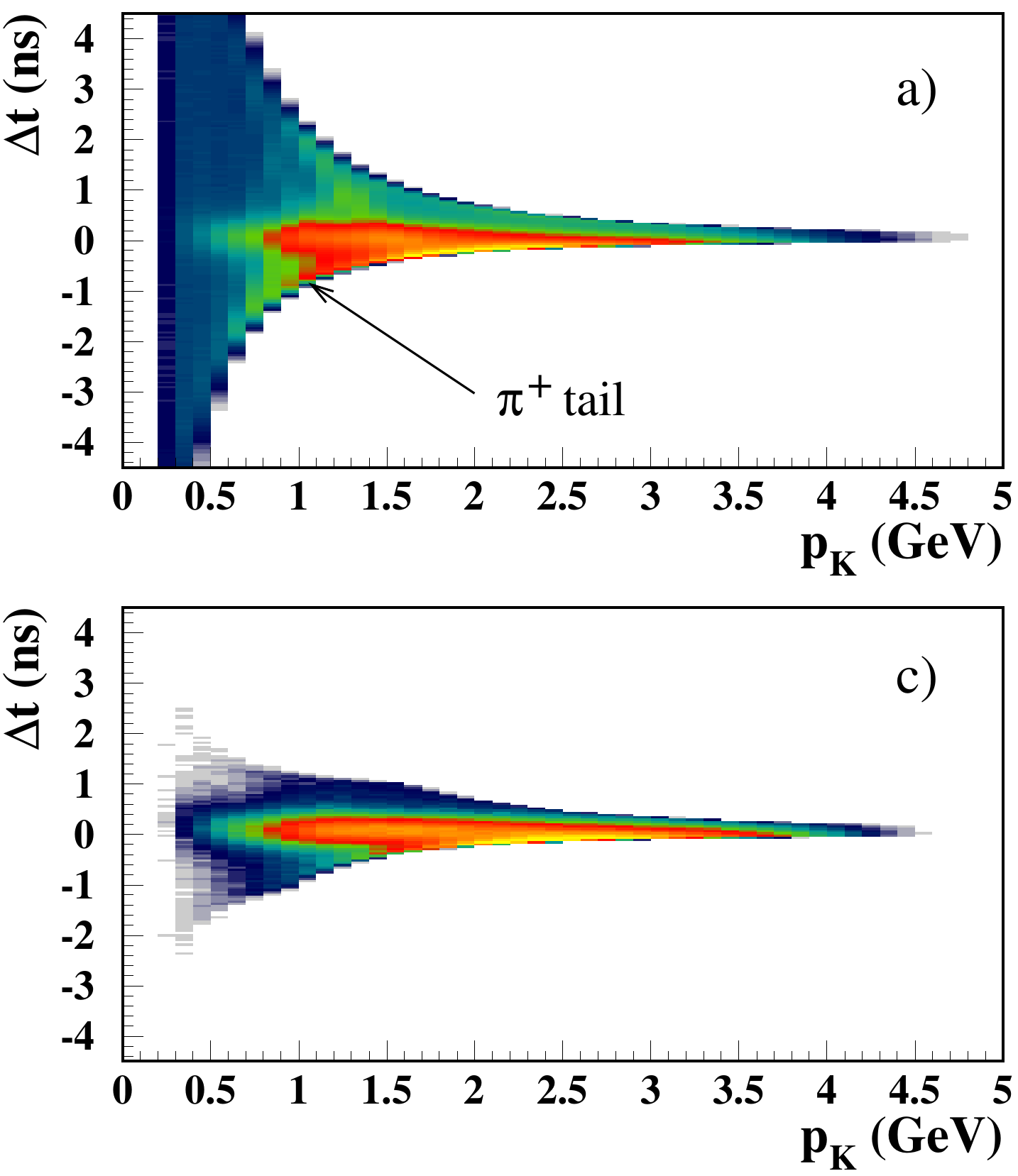}
  \includegraphics[height=7cm]{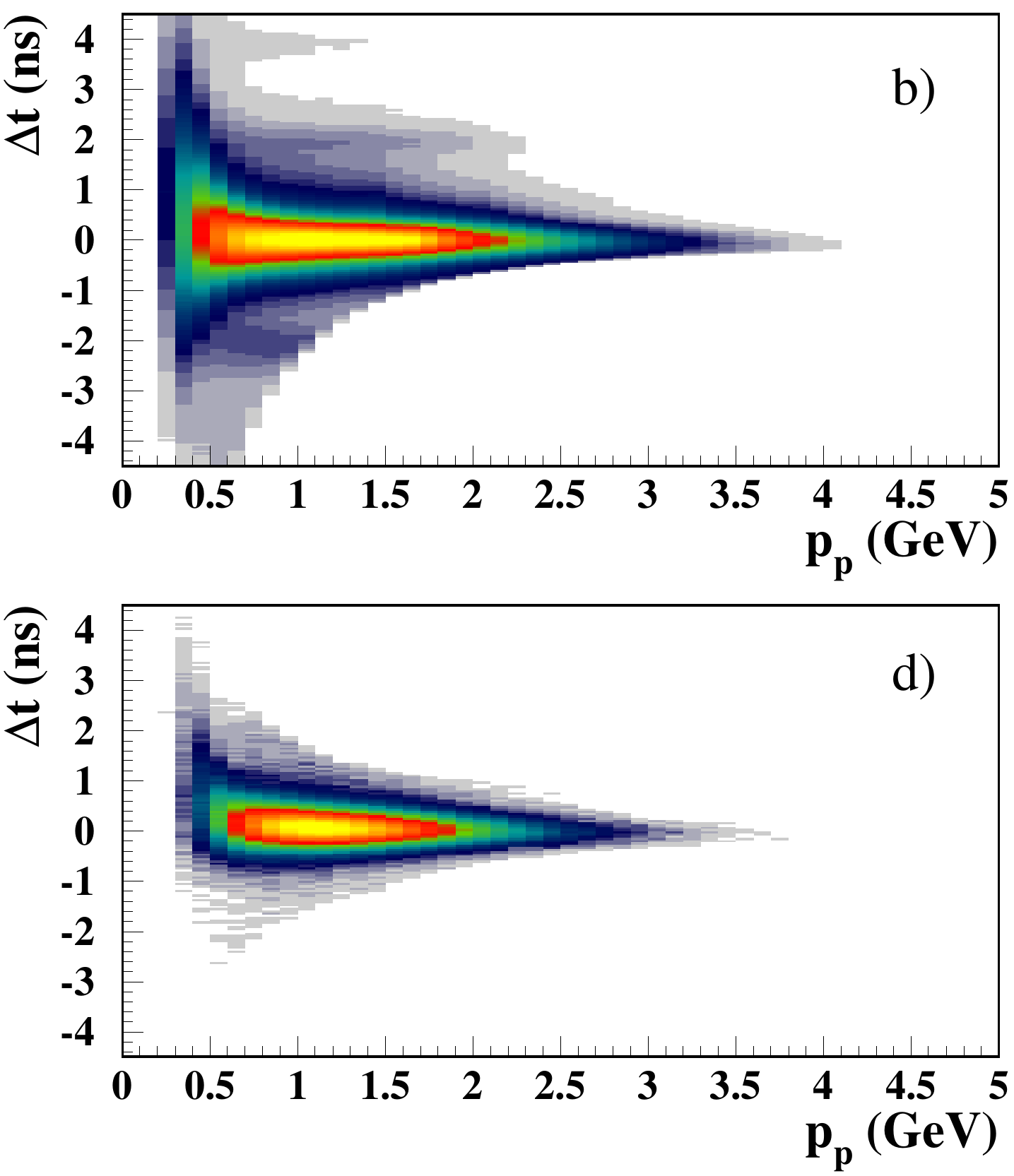}
  \caption{(Color online) Minimum $\Delta t$ (ns) vs.~$p$ (GeV) distributions
    for identified kaons (left) and protons (right).  Plots a) and b) show the distributions without
    any cuts. Plots c) and d) show the same distributions for kaons and protons after
    applying the $\Lambda$ missing-mass and the $\pi$ missing-mass-squared cuts shown in
    Fig.~\ref{mm_l0_corel}.}
  \label{h-id}
\end{figure*}

For all positive tracks, $\Delta t$ was calculated three times for assumed
particle masses of a pion, kaon, and proton.  The mass that gave the smallest
$\Delta t$ was assigned to the hadron. Figs.~\ref{h-id}a and \ref{h-id}b show the
minimum $\Delta t$ vs.~$p$ distributions for identified kaons and protons before
any cuts, respectively. The faint horizontal bands at $\pm$2~ns and $\pm$4~ns in
Fig.~\ref{h-id}b are due to accidental tracks from different beam bunches of
the accelerator. Figs.~\ref{h-id}c and \ref{h-id}d show the same distributions for
kaons and protons, respectively, after applying the $\Lambda$ missing-mass and
$\pi$ missing-mass-squared cuts described in the next paragraph. The application
of these cuts effectively removed the accidental coincidences and most of the
background in the kaon distribution, which consisted of pions and protons
misidentified as kaons. As $\beta \to 1$, the pion, kaon, and proton bands
started to overlap, leading to a background that was subtracted later in the
analysis (see Sec.~\ref{background}).

\begin{figure*}[hbtp]
  \includegraphics[height=7cm]{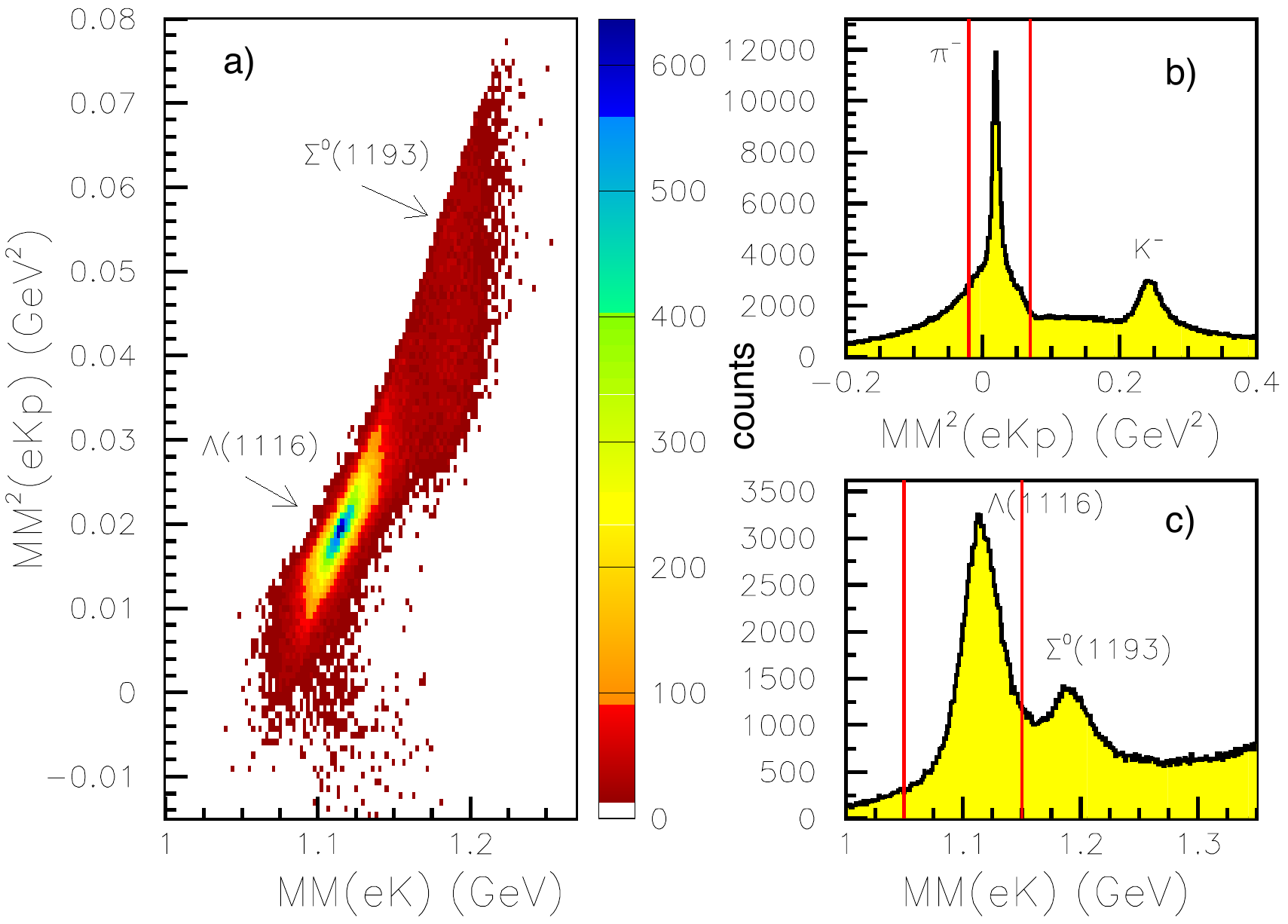}
  \caption{(Color online) a) Reconstructed missing mass squared $MM^2(e'K^+p)$ (GeV$^2$) vs. baryon
    missing mass $MM(e'K^+)$ (GeV). b) Missing mass squared distribution $MM^2(e'K^+p)$ 
    (GeV$^2$). The red lines show the applied cut, which includes events with
    only a missing pion ($\Lambda$ events) and events with a missing pion
    plus a photon ($\Sigma^0$ events). Negative values are due to finite resolution effects. 
    c) Hyperon missing mass $MM(e'K^+)$ distribution
    after applying the $\pi$ missing-mass-squared cut. The red lines in this plot
    show the missing mass range over which the background-subtracted yields are
    integrated for the final $\Lambda$ sample selection. All plots require a
    detected proton.}
  \label{mm_l0_corel}
\end{figure*}

The final-state hyperons were identified using the missing-mass
technique. The correlation of missing mass squared $MM^2 (e'K^+p)$ vs.~$MM(e'K^+)$ is shown in
Fig.~\ref{mm_l0_corel}a. Figs.~\ref{mm_l0_corel}b and \ref{mm_l0_corel}c are the projections
of the correlation plot onto the respective axes. Since protons will also be
present from higher mass hyperon decays, those events cannot be fully eliminated
from the $\Lambda$ mass peak. A cut was placed on the $MM^2(e'K^+p)$ missing mass
squared distribution from $-0.02$ to 0.07~GeV$^2$ (shown in red in
Fig.~\ref{mm_l0_corel}b). This cut was chosen so that events with either a missing
pion alone or a missing pion plus a photon remain, so that the full
$\Sigma^0$ peak was preserved in the hyperon mass distribution. The low-mass tail
of the $\Sigma^0$ peak beneath the $\Lambda$ was removed by a fitting procedure during background
subtraction and the $\Lambda$ yield was determined over the range shown by the
red lines in Fig.~\ref{mm_l0_corel}c.  The background subtraction procedure is
discussed in Section~\ref{background}.

\subsection{Data Binning}
\label{binning}
\begin{figure}[hbtp]
  \includegraphics[width=1\columnwidth]{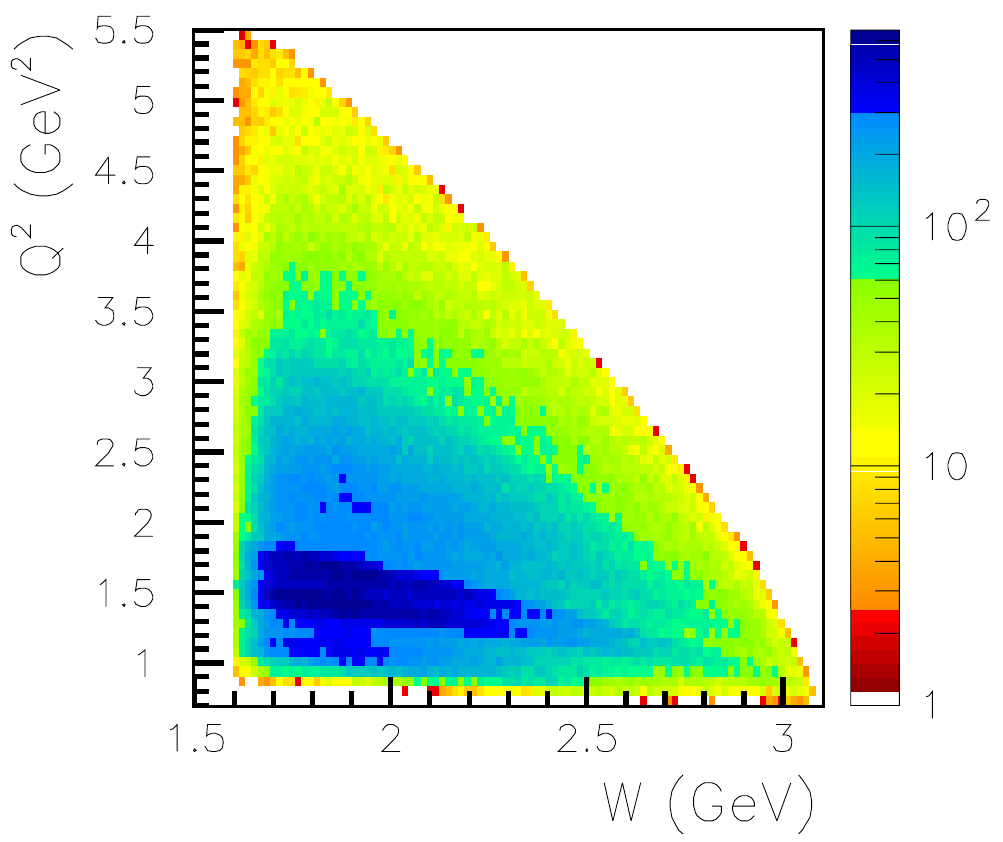}
  \caption {(Color online) Distribution of $K^+\Lambda$ events in $Q^2$ and $W$. The 
   ${\cal P}_N^0$ values are measured for $W$ up to 2.7~GeV only and summed over the full $Q^2$ range.}\label{fig:Q2W}
\end{figure}

We employed two different binning schemes for this work.  In Binning I shown in
Table~\ref{bins}, the data were binned in the invariant energy, $W$, and the cosine of the kaon 
production angle in the center-of-mass frame ($\theta_K^{CM}$), and were summed over $Q^2$
and $\Phi$. The bin widths were chosen to have approximately equal statistical uncertainties in each 
kinematic bin. In Binning II, also summed over $\Phi$, much larger bins in $W$ and $\cos\theta_K^{CM}$
were employed to study the $Q^2$ dependence of the
polarization. Fig.~\ref{fig:Q2W} shows the kinematic extent of the $K^+\Lambda$ data in terms of $Q^2$
versus $W$. The $Q^2$ range spanned by the data depends 
strongly on $W$.
\begin{table}[hbtp]
  \begin{ruledtabular}
    \begin{tabular}{|r|r|r|r|}
      \multicolumn{4}{|c|} {Binning I} \\ \hline
      Variable             &  Range      &   \# of bins  & Bin Width  \\ \hline 
      $\cos\theta_K^{CM} $  & (-1.0, 0.0) &  2            &    0.5     \\
                           & (0.0, 1.0)  &  5            &    0.2      \\ \hline
      $W$                  &  1.6-2.1~GeV &  20          &    25~MeV   \\
                           &  2.1-2.7~GeV &  12          &    50~MeV   \\ \hline
      \multicolumn{4}{|c|} {Binning II} \\ \hline
      $\cos\theta_K^{CM} $   & (-1.0, 0.0) &  1            &    1.        \\
                           & (0.0, 0.4)   &  1           &    0.4       \\
                           & (0.4, 0.8)   &  1           &    0.4       \\
                           & (0.8, 1.0)   &  1           &    0.2       \\ \hline
      $W$                  &  1.6-2.4~GeV &  4           &    200~MeV   \\ \hline
      $Q^2$                &  0.8-3.2~GeV$^2$ &  4        &    0.6~GeV$^2$   \\
    \end{tabular}
    \caption{Data binning for the induced $\Lambda$ polarization analysis.  Binning I is used to study the $W$ and
      $\cos\theta_K^{CM}$ dependencies and Binning II is used to study the $Q^2$
      dependence.}
    \label{bins}
  \end{ruledtabular}
\end{table}
%

\subsection{$\Lambda$ Polarization Extraction}
\label{qm_l0}
Although the $\Lambda$ is produced in a strong hadronization process
it can only decay weakly.  The main decay modes of the $\Lambda$ are $p \pi^-$
and $n \pi^0$ with branching ratios of 64\% and 36\%,
respectively~\cite{pdg}.  The $\Phi$-integrated proton angular distribution from the
$\Lambda$ decay in the $\Lambda$ rest frame is given by
\begin{equation}
  \label{dndcos}
  \frac{dN}{d\cos\theta_p^{RF}} =  N_0 (1+\alpha {\cal P}_j \cos \theta_p^{RF}),
\end{equation}
where ${\cal P}_j$ is the $\Phi$-integrated polarization of the $\Lambda$ for an arbitrary coordinate 
($\hat{t}$, $\hat{n}$, $\hat{l}$) in the $\Lambda$ rest frame, $\theta_p^{RF}$ is the  angle of the decay proton relative to
the respective $\hat{t}$, $\hat{n}$, or $\hat{l}$ axis in the $\Lambda$ rest frame, and $\alpha=0.642 \pm 0.013$~\cite{pdg} 
is the weak decay asymmetry parameter. The uncertainty in $\alpha$ is 
treated as a systematic uncertainty in the final result.

The induced polarization for a given coordinate can be extracted by forming the forward-backward yield
asymmetry with respect to $\cos\theta_p^{RF}=0$. Integrating
Eq.~\ref{dndcos} from 0 to 1 (forward) and -1 to 0 (backward), gives the corresponding yields
$N^+$ and $N^-$ as:
\begin{eqnarray}
  \label{np_nm}
  N^+ &=& \int_0^1 N_0 (1+\alpha {\cal P}_j^0 \cos \theta_p^{RF})d\cos \theta_p^{RF}
    \nonumber \\ 
  &=& N_0+N_0\frac{\alpha {\cal P}_j^0}{2} \nonumber \\
  N^- &=& \int_{-1}^0 N_0 (1+\alpha {\cal P}_j \cos \theta_p^{RF})d\cos \theta_p^{RF}
    \nonumber \\ 
  &=& N_0-N_0\frac{\alpha {\cal P}_j^0}{2}.
\end{eqnarray}
The forward-backward yield asymmetry with respect to a given axis $j=(\hat{t},\hat{n},\hat{l})$, 
$A_j$, is then defined as
\begin{equation}
\label{fb_asym}
  A_j = \frac{N^+ - N^-}{N^+ + N^-} = \frac{\alpha {\cal P}_j^0}{2}, 
\end{equation}
and the induced polarization can be expressed in terms of the asymmetry as
\begin{equation}
\label{fb_asym_pol}
  {\cal P}_j^0 = \frac{2A_j}{\alpha}= \frac{2}{\alpha}\cdot \frac{N^+ - N^-}{N^+ + N^-}.
\end{equation}

\subsection{Background Subtraction}
\label{background}

In order to form the forward-backward yield asymmetries, the background-subtracted
$\Lambda$ yields must be determined from the $e'K^+$ missing mass distributions.
To determine the $\Lambda$ yields, the contributions of background beneath the
$\Lambda$ peak had to be accounted for. This included the background both from the 
low-mass tail of the $\Sigma^0$ peak and from hadron misidentification. The respective 
yields of $\Lambda$, $\Sigma^0$, and other background were determined by a fit to the
missing mass distributions for each kinematic bin.  The $\Lambda$ and $\Sigma^0$ peaks were fit with
functional forms, $f_\Lambda$ and $f_\Sigma$, respectively, that were motivated
by the results of a Monte Carlo simulation that was well matched to the data. The simulation suggested that the
line shape for each of the hyperons was well represented by a Gaussian plus two Lorentzians for the high and low
mass sides of the peaks. This form accounted for the finite detector resolution, as
well as the radiative tail on the high mass side of the peak. The background
beneath the hyperons was primarily from pions misidentified as kaons and depended
strongly on kinematics. To define the shape of this background, $f_{BG}$,
templates were generated from the data by intentionally misidentifying pions as
kaons. The scale for the background template was allowed to float as a free
parameter in the fitting procedure.

The total fit function was then defined as
\begin{equation}
  \label{eq:totfit}
  F_{TOTAL} = f_\Lambda+f_\Sigma+f_{BG}, 
\end{equation}
with 
\begin{eqnarray}
  f_\Lambda &=& G_\Lambda+L_\Lambda^L+L_\Lambda^R, \nonumber \\
  f_\Sigma &=& G_\Sigma+L_\Sigma^L+L_\Sigma^R,  \nonumber \\
  f_{BG} &=& C_{BG}\times (background\ template),  \nonumber 
\end{eqnarray}
in which $G$, $L^L$, and $L^R$ are the Gaussian and the left and right Lorentzian
functions (low and high mass sides), respectively, and  
$C_{BG}$ is the amplitude parameter for the background from hadron misidentification.

\begin{figure*}[hbt]
 
     \includegraphics[height=4.cm,width=6.cm]{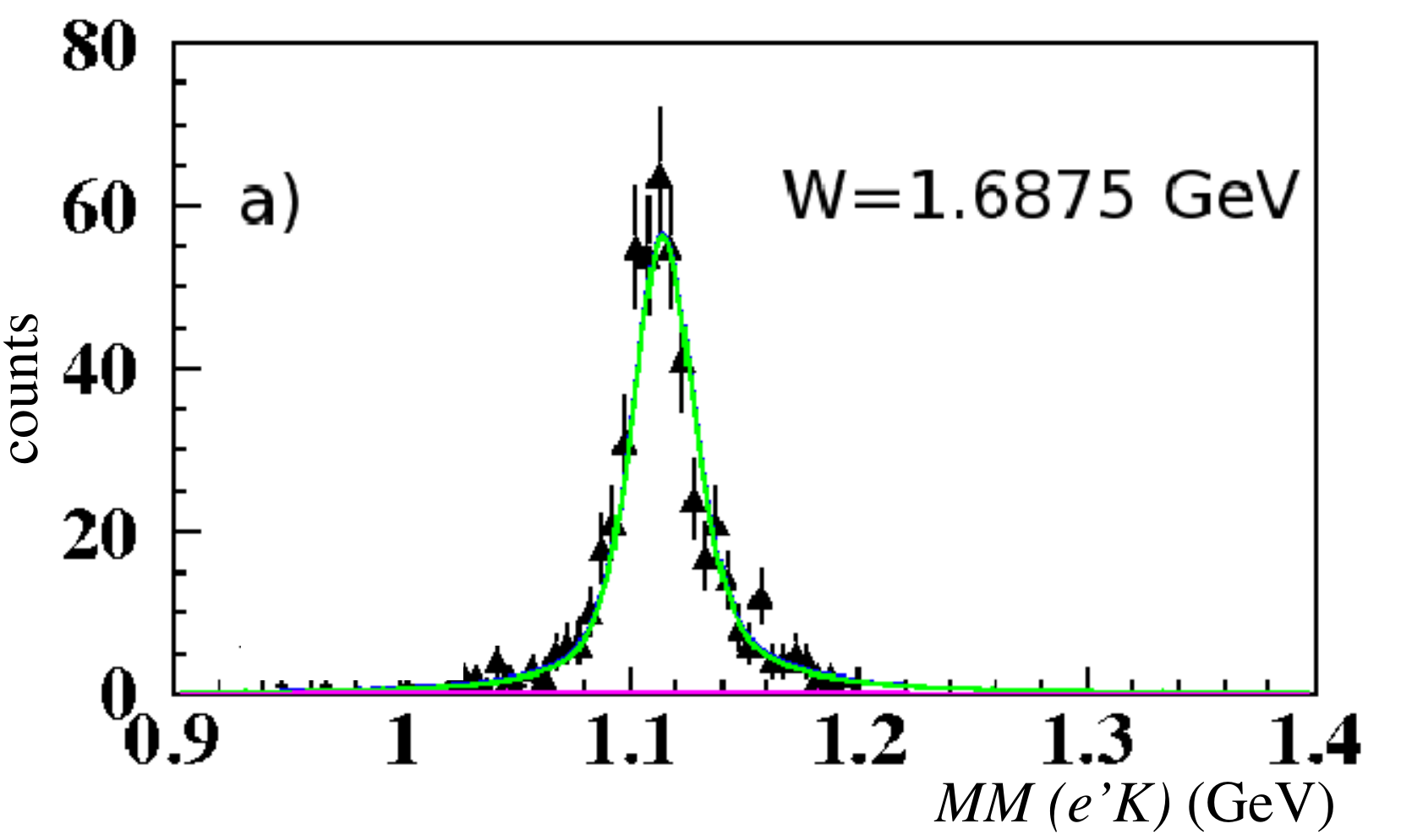}
     \includegraphics[height=4.cm,width=6.cm]{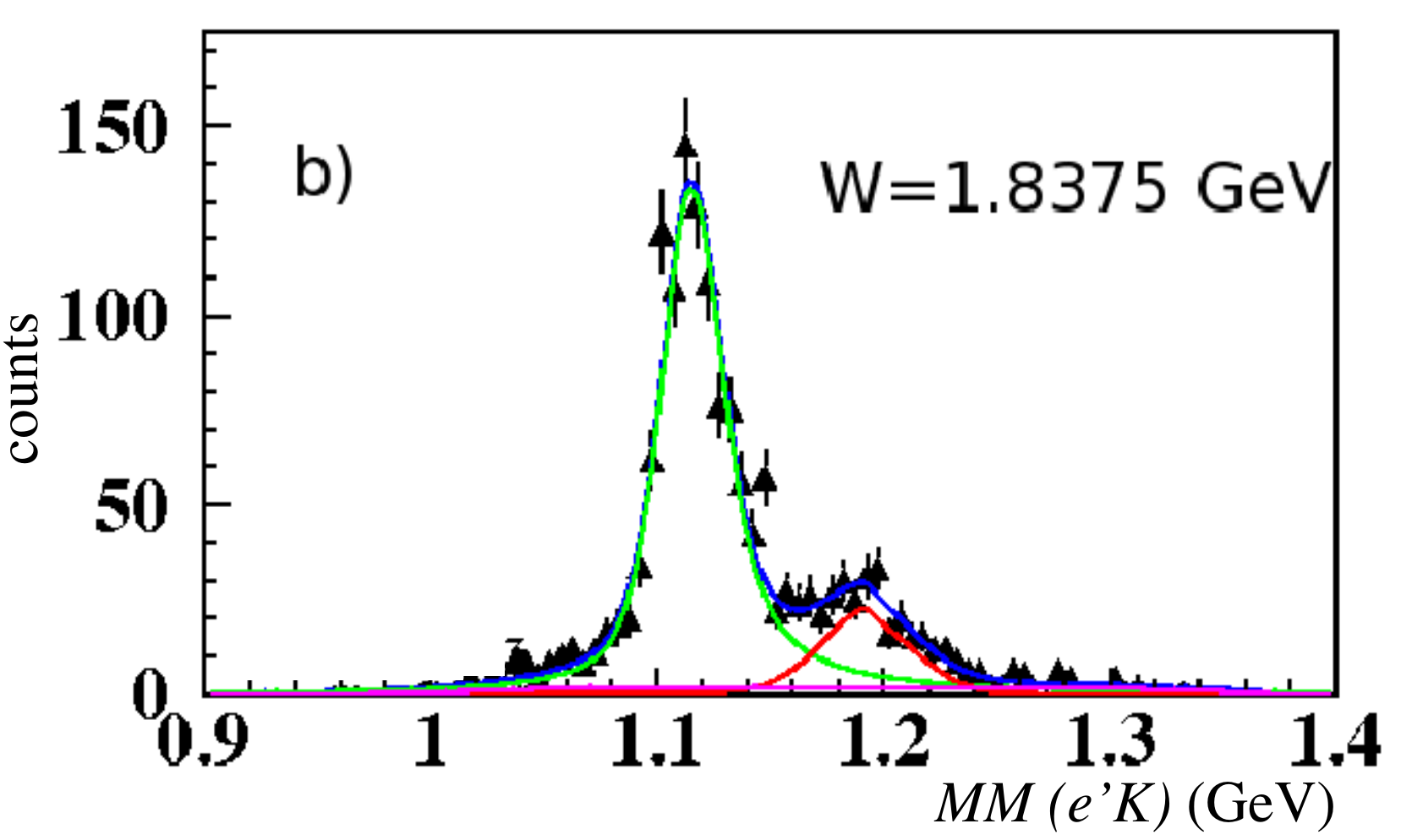}
     \includegraphics[height=4.cm,width=6.cm]{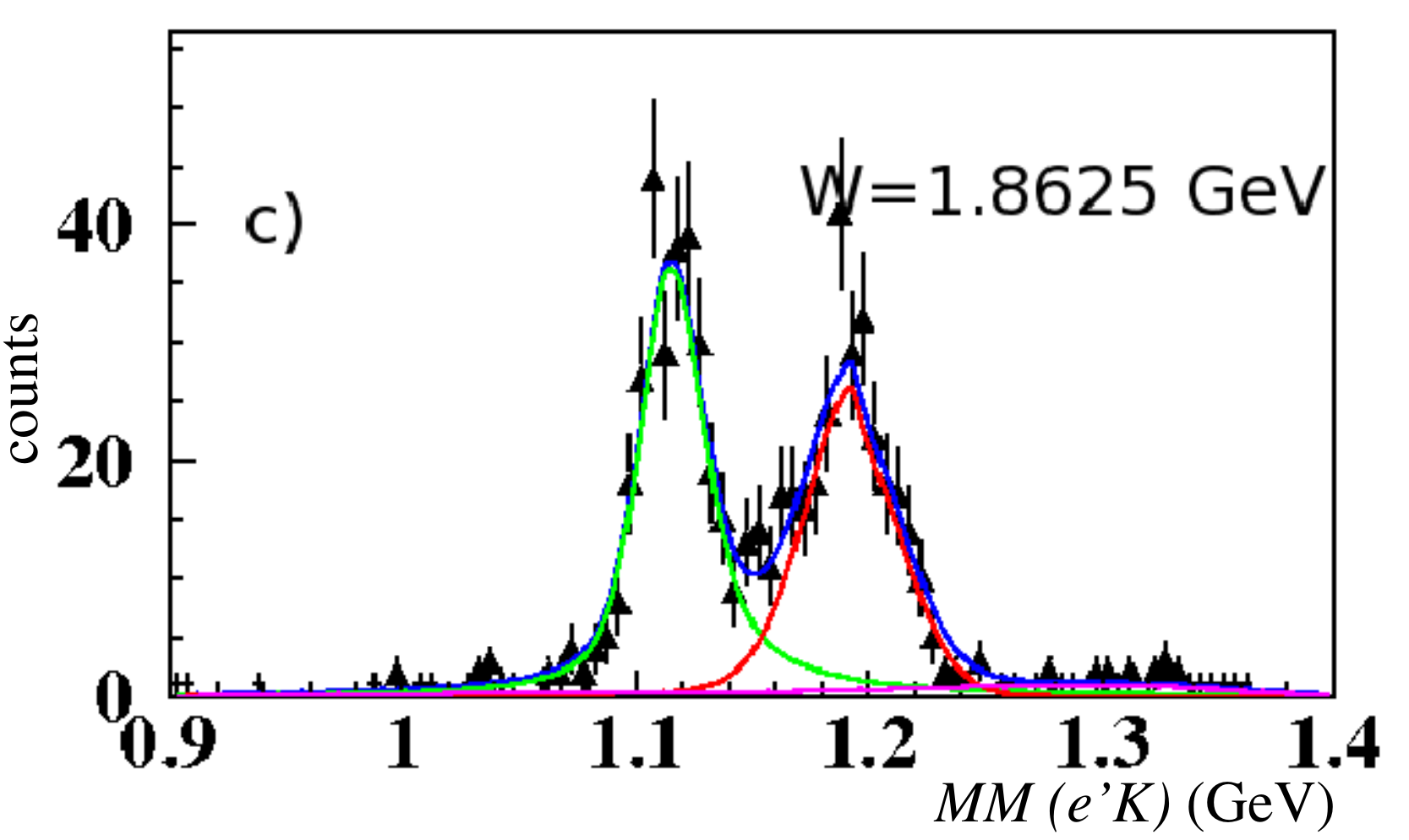}
     \includegraphics[height=4.cm,width=6.cm]{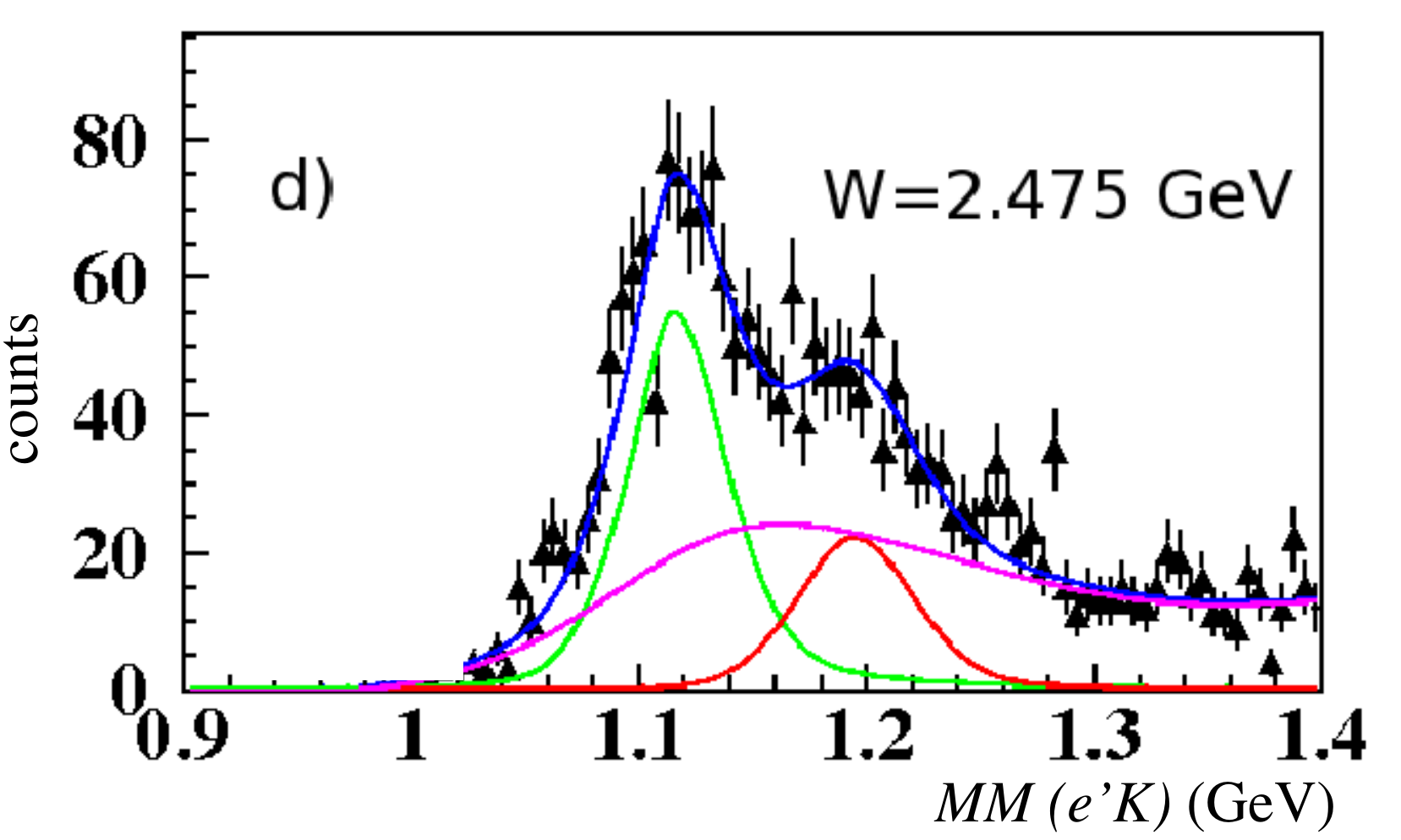}
     \caption{(Color online) Typical fits to the reconstructed hyperon mass
       spectra for different $W$ bins summed over $Q^2$ using Eq.~\ref{eq:totfit}.  Panels a, b, c, and d correspond to 
       $\cos\theta_K^{CM}=0.5$, 0.9, 0.1, and 0.9, respectively.
       The green curves correspond to the $\Lambda$ peak, the red curves to $\Sigma^0$ peak, the magenta
       curves to the hadron-misidentification background, and the blue curves to the
       total fit function.} 
   \label{bkg_fit}
\end{figure*}

Several constraints were applied to the fits.  The high and low-mass
Lorentzians were required to have the same magnitude at the peak of the hyperon.  The
centroids of the Gaussians and Lorentzians were restricted to the PDG values of the
hyperon masses \cite{pdg}. To further constrain the fits, we assumed that the
shape parameters (Gaussian and Lorentzian widths) must vary smoothly
from one kinematic bin to the next and that the shape of the peaks must be 
the same for forward-going and backward-going protons in the hyperon
rest frame. We parameterized the shape parameters as a function of
$W$, thereby reducing the number of free parameters in the final fit to
six for each of the backward and forward yields, where only the
Gaussian and Lorentzian amplitudes were allowed to vary freely.
Typical sample fits are shown in Fig.~\ref{bkg_fit}. The average $\chi^2$ per degree
of freedom, $\chi^2/ndf$, was approximately 1.

The number of $\Lambda$'s in each kinematic bin, corresponding to forward and backward-going 
protons, was determined by integrating the functions corresponding to the $\Lambda$ peak within
the missing mass range from 1.05~GeV to 1.15~GeV.  The
background-subtracted yields, $Y_{\pm}$, have a statistical 
uncertainty given by 

\begin{equation}
 \label{yielderr}
 \delta Y_{\pm}^2=\displaystyle\sum\limits_{i}^{n}\displaystyle\sum\limits_{j}^{n}
 \frac{\partial f_\Lambda} 
 {\partial a_i}\frac{\partial f_\Lambda}{\partial a_j}\epsilon_{ij},
\end{equation} 
where $n$ is the number of free parameters, $\epsilon_{ij}$ is the correlation
matrix from the fit, and $a_i$ and $a_j$ are the fit parameters, and  $f_\Lambda$ is
the $\Lambda$ peak fit function integrated within the missing mass range.

\subsection{Acceptance Corrections}
\label{acc}

The final background-subtracted forward/backward yields were corrected
for acceptance and efficiency effects using a GEANT-based Monte Carlo simulation. In the first stage of the
simulation, $ep \to e'K^+\Lambda$ events were generated with a
$t$-slope-modified phase-space generator. The event generator scaled the phase
space cross section by a factor of $e^{-bt}$, where $b$ is the $t$-slope
parameter, and the Mandelstam variable $t=(k_\gamma-p_K)^2$ is the square of the difference between the
virtual photon and kaon four momenta. The choice of $b$=0.3~GeV$^{-2}$ yielded a distribution that was a reasonable 
match to the data. To first order, the electron and kaon acceptances cancel out in the forward-backward asymmetry so a
perfect match of the angular distributions of kaons and electrons between data and Monte 
Carlo is not necessary. The generated events were then processed with the GSIM package, which
is the GEANT simulation of the CLAS detector. Although only the external final-state
radiative effects were included in GSIM, any electron or kaon radiative effects are the same for forward and backward going
protons from the decay of the $\Lambda$ and will cancel out in the forward-backward asymmetry.

In the first iteration of the correction, the induced polarization was assumed
to be zero, leading to a uniform proton distribution in $\theta_p^{RF}$.  The
particles were then propagated through CLAS and the detector response was
recorded in the same way as for the experimental data. The GSIM simulation
assumed a perfect detector system, so the known inefficiencies and the
resolutions of the different detector components were taken into account in the
next step by the GSIM post-processing (GPP) package, which smeared the DC and TOF
resolutions to match the experimental data. The simulated data were processed
identically to the experimental data.

The acceptance factors in this analysis were defined as the ratio of the
reconstructed events to the generated events in the same kinematic bin.  Two
acceptance factors $f_{\pm}$ were defined in each kinematic bin corresponding to
forward and backward-going protons with respect to a given spin quantization
axis in the $\Lambda$ rest frame and are given by
\begin{equation}
 \label{facc}
  f_{\pm}=\frac{N^{\pm}_{Detected}}{N^{\pm}_{Thrown}}.
\end{equation} 
The numerator, $N^{\pm}_{Detected}$, is the number of detected $\Lambda$s after
all cuts were applied and $N^{\pm}_{Thrown}$ is the number of generated
events. 
%

In the second iteration of the acceptance correction procedure, the induced
polarization results determined using the correction factors of the zero-polarization $\Lambda$-decay simulation were
then used as the input polarization of the simulated data. This gave a more
realistic decay-proton distribution. The acceptance factors were then
recalculated.

\begin{figure}[bt]
   \includegraphics[height=8.5cm,width=8cm]{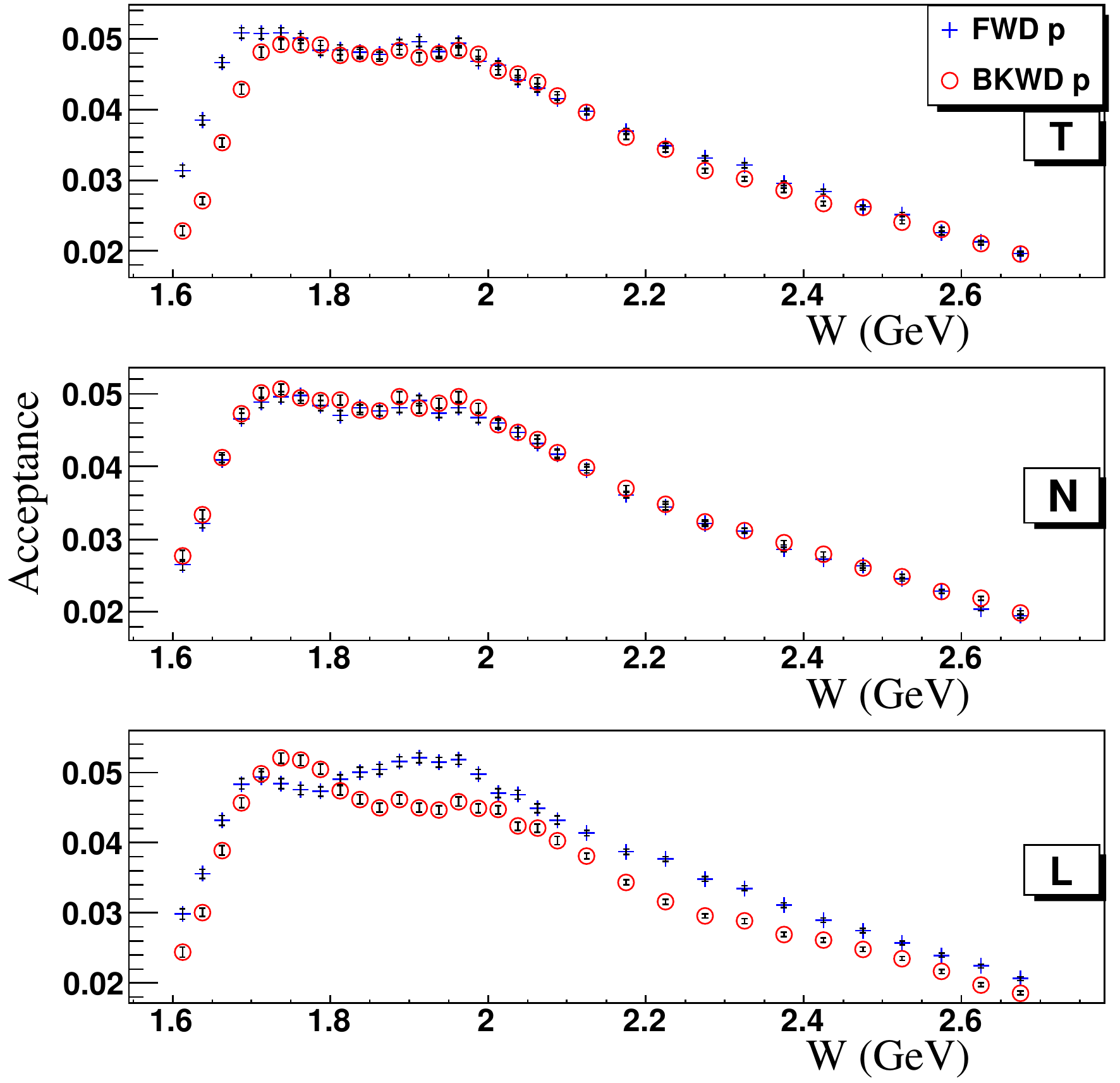}
   \caption[]{(Color online) Dependence of the acceptance factors on $W$
     for forward-going (blue crosses) and backward-going (red circles) protons at
     0.8$<\cos\theta_K^{CM}<$1 with respect to the $\hat{t}$, $\hat{n}$, and $\hat{l}$
     polarization axes.}
 \label{facc_vs_w}
\end{figure}

The $W$ dependence of the acceptance factors are plotted in Fig.~\ref{facc_vs_w}
for the most forward kaon angular bin. As can be seen from the plot, the normal
component of the polarization has nearly identical acceptances for both forward-
and backward-going protons, while the other two components have some rather
large differences in the forward and backward acceptances and are therefore more
sensitive to acceptance effects. This statement is true for all $\cos\theta_K^{CM}$
angles. As previously mentioned, the $\hat{t}$ and $\hat{\ell}$ components (see
Fig.~\ref{fig-kin}) of the induced polarization must vanish when integrated over
$\Phi$, which will only happen if the acceptance factors for these components
are properly accounted for (see Sec.~\ref{sec:systematics}).

The acceptance-corrected yields for the forward and backward
directions are given by
\begin{equation}
 \label{corr_yield}
 N_{\pm}= Y_{\pm}/f_{\pm}.
 \end{equation} 
Here, $Y_{\pm}$ is the background-subtracted, uncorrected yield, obtained by fitting as  described in the 
previous section and $f_{\pm}$ is the acceptance correction factor from Eq.~\ref{facc} after applying the second iteration of acceptance
corrections. 

\section{Systematic Uncertainties}
\label{sec:systematics}

There were four primary sources of point-to-point systematic uncertainties that
we identified in this analysis.  These sources were from background
subtraction, acceptance corrections, geometrical fiducial cuts, and hyperon peak
fitting. The systematic uncertainties were determined for each source by
comparing the nominal polarization results in each $\cos \theta_K^{CM}$ kinematic
bin to the results obtained with alternative cuts or corrections. The systematic
uncertainties were estimated as the uncertainty-weighted average polarization
difference defined by:
\begin{eqnarray}
 \label{avg_P}
 \delta P &=& \sqrt \frac{\displaystyle\sum\limits_{i=1}^n
   \frac{[P_i^{nom}-P_i^{alt}]^2}{(\delta P_i^{nom})^2}} 
 {\displaystyle\sum\limits_{i=1}^n \frac{1}{(\delta P_i^{nom})^2}}.
\end{eqnarray}
Here the summation goes over all $W$ points for each $\cos
\theta_K^{CM}$ bin. We found that within a given
$\cos\theta_K^{CM}$ bin, variations with $W$ were statistically
distributed. 
\begin{table*}[t]
\begin{center}
 \begin{tabular}{|r|r|r|r|r|r|r|r|r|}
  \hline
  \multicolumn{8}{|c|}{Systematic Uncertainties} \\ \hline
  \backslashbox{Source}{$\cos \theta_K^{CM}$}  & (-1.0,-0.5) &
  (-0.5,0.0) & (0.0,0.2) & (0.2,0.4) & (0.4,0.6) & (0.6,0.8) &
  (0.8,1.0) \\ \hline
  Background subtraction & 0.042 & 0.025 & 0.047 & 0.036 & 0.046 & 0.041 & 0.033 \\ \hline
  Acceptance corrections & 0.082 & 0.074 & 0.072 & 0.080 & 0.079 & 0.069 & 0.063 \\ \hline 
  Geometrical            & 0.051 & 0.030 & 0.031 & 0.025 & 0.015 & 0.032  & 0.032 \\
  Fiducial Cut           &       &       &       &       &       &       &       \\ \hline
  Fitting                & 0.056 & 0.058 & 0.056 & 0.041 & 0.034 & 0.032 & 0.034 \\ \hline 
  $\delta\alpha/\alpha$  & 0.020 & 0.020 & 0.020 & 0.020 & 0.020 & 0.020 & 0.020 \\ \hline
  Total point-to-point   & 0.120 & 0.103 & 0.108 & 0.101 & 0.100 & 0.093 & 0.086 \\ \hline 
 \end{tabular}
 \caption{Estimated systematic uncertainties. The total point-to-point
   systematic uncertainty on the measured $\Lambda$ induced polarization (last row) is the 
   quadrature sum of the individual contributions.}  
 \label{sys_err_table}
\end{center}
\end{table*}

The systematic uncertainties from all sources are detailed in
Table~\ref{sys_err_table}. Uncertainties associated with the background
subtraction were determined by widening the cut on $MM^2(e'K^+p)$, thus letting
in more background. The estimated background-related uncertainty is between
0.025 and 0.047. The acceptance correction uncertainty was determined by varying
the $t$-slope of the event generator over the range 0.1 to 1.0~GeV$^{-2}$. The
estimated acceptance-correction uncertainty was between 0.063 and 0.082. The
geometrical fiducial cuts on the proton acceptance were varied between tighter and 
looser cuts over a sensible range, leading to an
estimated systematic uncertainty between 0.015 and 0.051. We tested different
methods of fitting the hyperon spectrum including using different types of
fitting routines and allowing shape parameters to float freely as opposed to
using smoothly varying shape parameters. The estimated fitting uncertainty
varied between 0.032 and 0.058.

We generally see that the systematic uncertainties get bigger at larger kaon angles.  These bins have 
the largest statistical uncertainties and therefore estimating systematic uncertainties becomes less certain. 
The overall systematic uncertainty, formed from a quadrature sum of the first four sources listed in
Table~\ref{sys_err_table}, varies between 0.086 in
the most forward kaon-angle bin to 0.120 at the most backward kaon-angle bin.

\begin{figure}[bt]
 \centering\includegraphics[height=6.5cm]{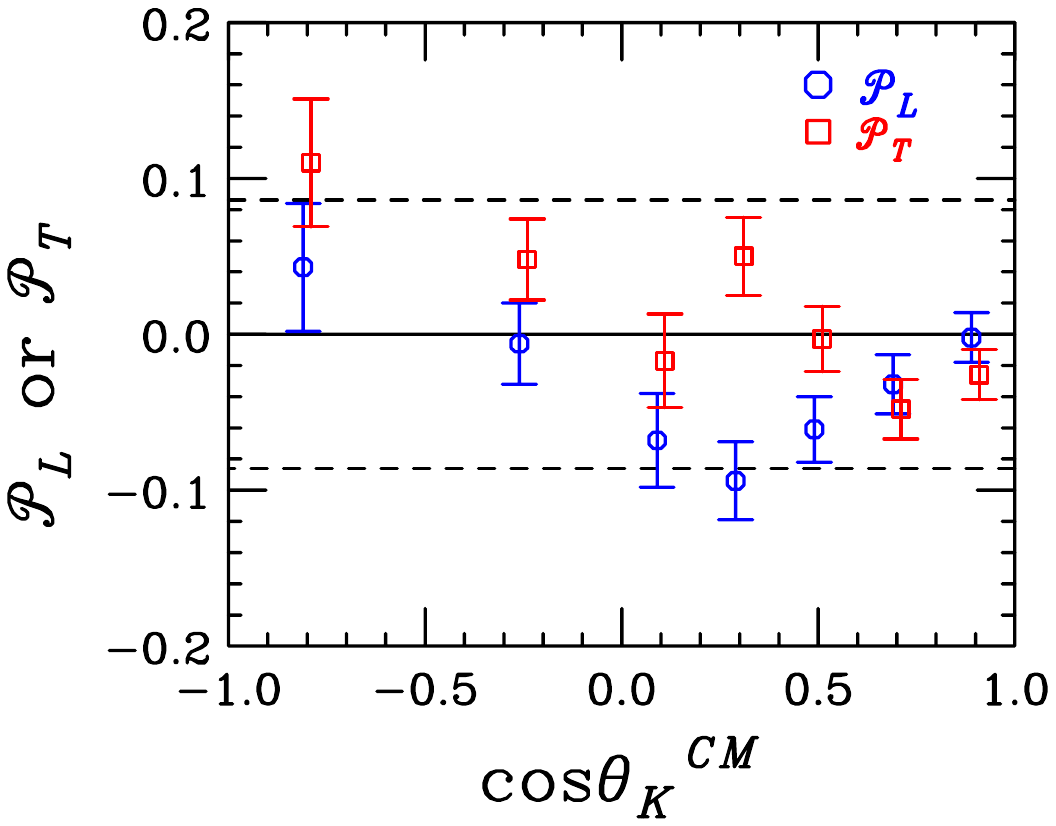}
 \caption{(Color online) $W$ averaged ${\cal P}_L^0$ (blue circles) and ${\cal P}_T^0$
   (red squares) values vs.~$\cos \theta_K^{CM}$ summed over $Q^2$. The dashed lines
   represent the lowest total point-to-point systematic uncertainty
   from Table~\ref{sys_err_table}.} 
 \label{plt_cos_p7}
\end{figure}

A powerful check of our systematic uncertainties was the measurement of the
$\Phi$-integrated longitudinal and transverse induced polarization components, 
${\cal P}_L^0$ and ${\cal P}_T^0$. Both of these quantities should be zero according to
Eq.~\ref{eqn-Pxp0_phi}. The $W$-averaged deviations from zero along with their
uncertainties are plotted vs.  $\cos \theta_K^{CM}$ in Fig.~\ref{plt_cos_p7} for
both ${\cal P}_L^0$ and ${\cal P}_T^0$. Within statistical uncertainties all ${\cal
  P}_L^0$ and ${\cal P}_T^0$ fall within the range of our smallest point-to-point
systematic uncertainty given by the dashed lines in the figure.

Finally, there is a scale-type uncertainty from the uncertainty on the $\Lambda$
decay parameter, $\alpha$.  This relative uncertainty is 0.020.

\section{Results and Discussion}
\label{sec:results}

\subsection{$Q^2$ Dependence} 
\label{q2_study}

\begin{figure*}[hbtp]

  \includegraphics[width=\textwidth]{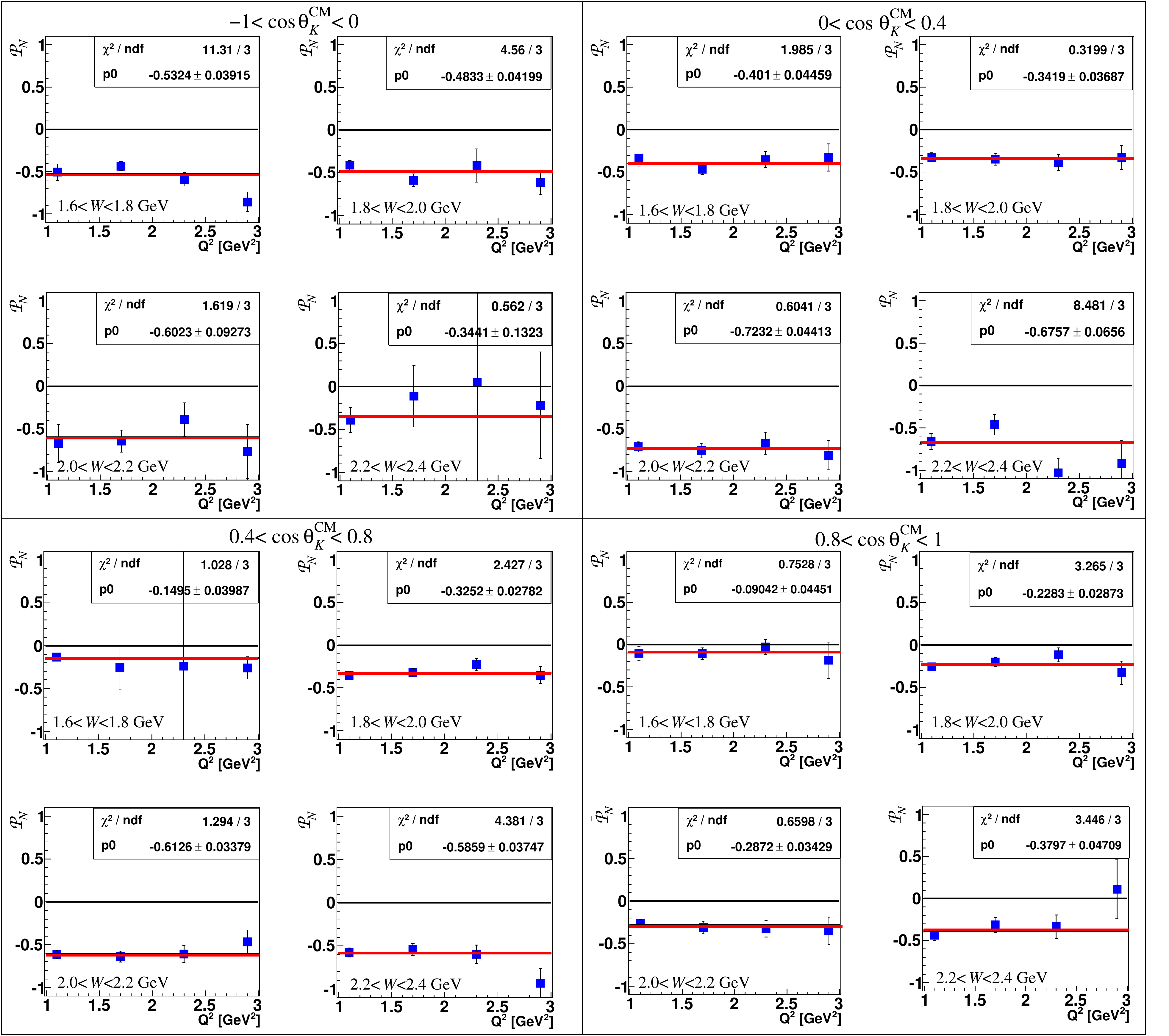}
  \caption[]{(Color online) Induced $\Lambda$ polarization ${\cal P}_N^0$ vs. $Q^2$ for different $\cos\theta_K^{CM}$
    and $W$ bins. The solid red lines are fits to a constant, and the error bars are statistical only. The results show 
    no significant dependence on $Q^2$  within our statistical uncertainties.}
  \label{pn_q2_cos1}
\end{figure*}

Fig.~\ref{pn_q2_cos1} shows the induced $\Lambda$ polarization ${\cal P}_N^0$ vs.~$Q^2$.  The data show a 
flat $Q^2$ dependence indicated by the quality of the constant fits.
The largest deviation from a flat distribution is in the bin for
$-1<\cos\theta_K^{CM}< 0$ and $1.6< W< 1.8$~GeV, which has a
$\chi^2/ndf=3.8$ and is essentially driven by a single data point. We will
discuss implications of the $Q^2$ dependence later. We took advantage of this
flat behavior and summed over $Q^2$ in order to improve the statistical
precision of the data for the $W$ and $\cos\theta_K^{CM}$ study.

%
%

\subsection{$W$ and $\cos\theta_K^{CM}$ Dependence}

\begin{figure*}[tp]
  \centering\includegraphics[height=19cm]{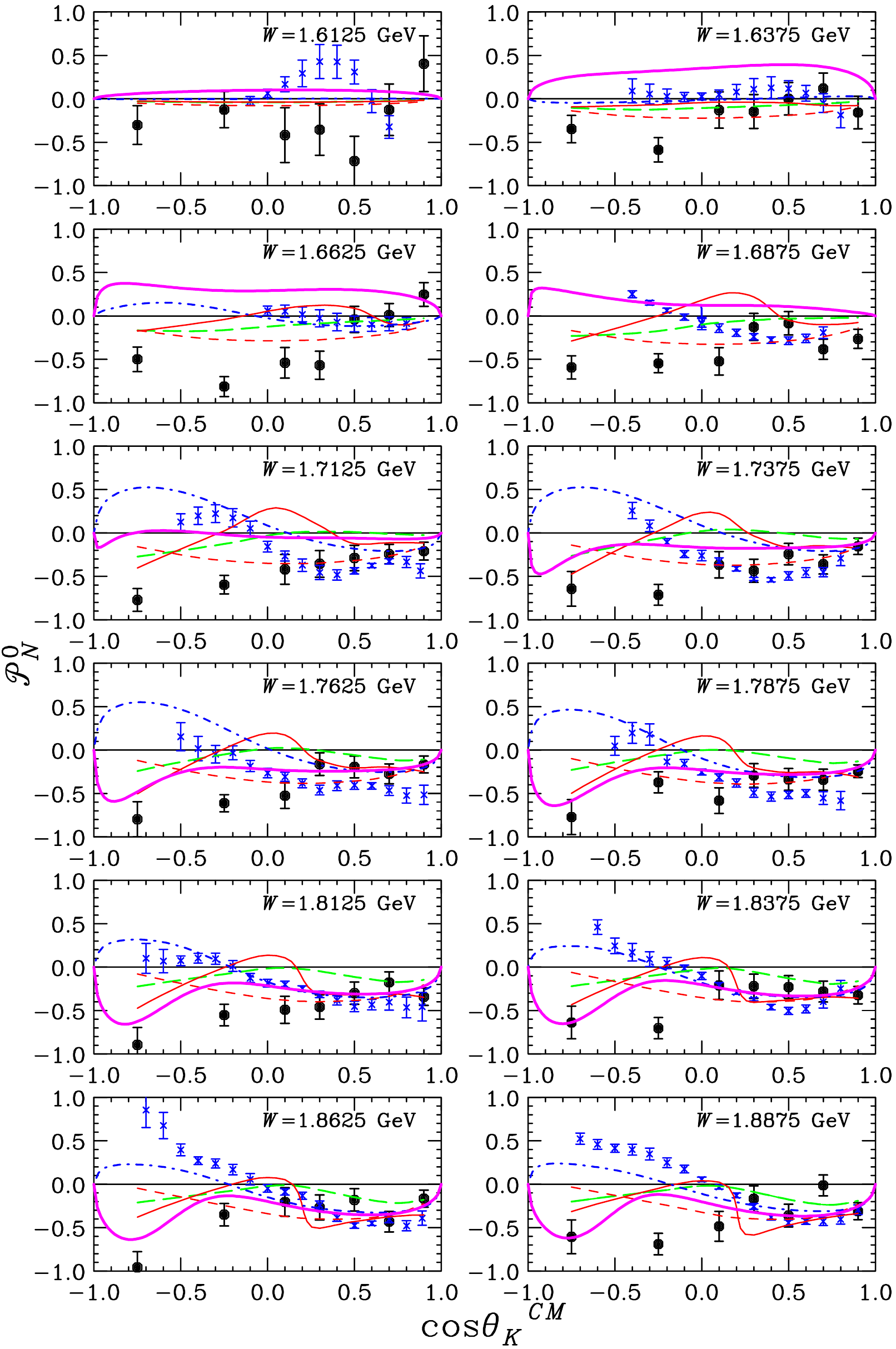}
  \caption[]{(Color online) Induced $\Lambda$ polarization ${\cal P}_N^0$
   vs.~$\cos \theta_K^{CM}$ for $W$ from 1.6125 to 1.8875~GeV at an average
   $Q^2$ of 1.90~GeV$^2$. The black circles are the results of this analysis and
   the blue crosses are the CLAS photoproduction results from
   Ref.~\cite{mccracken}. All data points show statistical uncertainties only. 
   The overlaid curves correspond to
   RPR-2007~\cite{corthals} (green long dash), RPR-2011~\cite{tom-vrancx}
   (red solid), RPR-2011NoRes~\cite{tom-vrancxPC} (red short dash), extended Kaon-Maid~\cite{Mart14} (blue dot-dash), and
   Maxwell~\cite{maxwell:12,maxwell:12-2} (magenta thick solid) model
   predictions, respectively.}
 \label{Pnvscos1}
\end{figure*}

\begin{figure*}[tp]
 \centering\includegraphics[height=19cm]{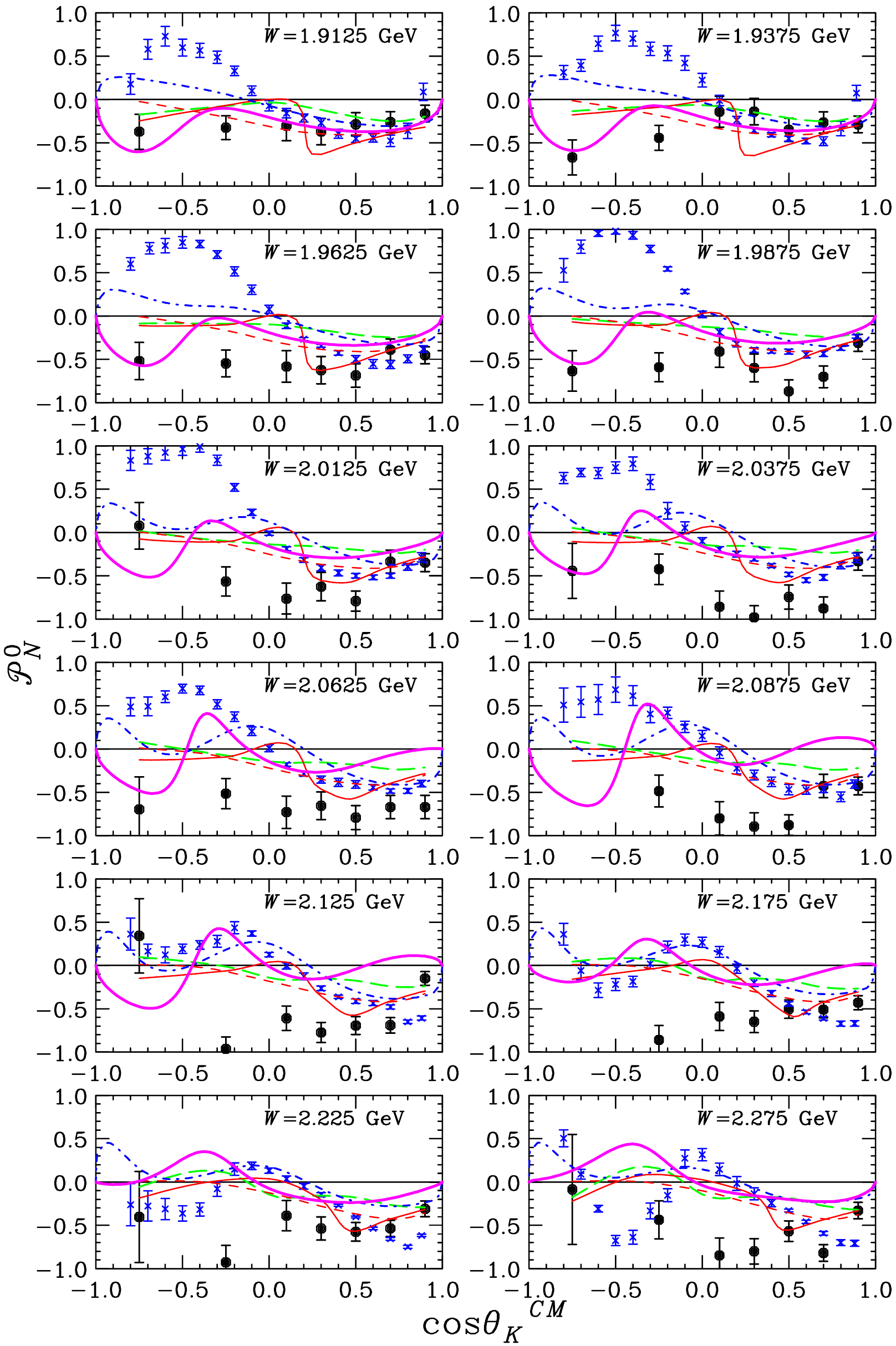}
 \caption[]{(Color online) Same as Fig.~\ref{Pnvscos1} except for $W$ from 1.9125 to 2.275~GeV.}
 \label{Pnvscos2}
\end{figure*}

\begin{figure*}[tp]
 \centering\includegraphics[height=12.89cm]{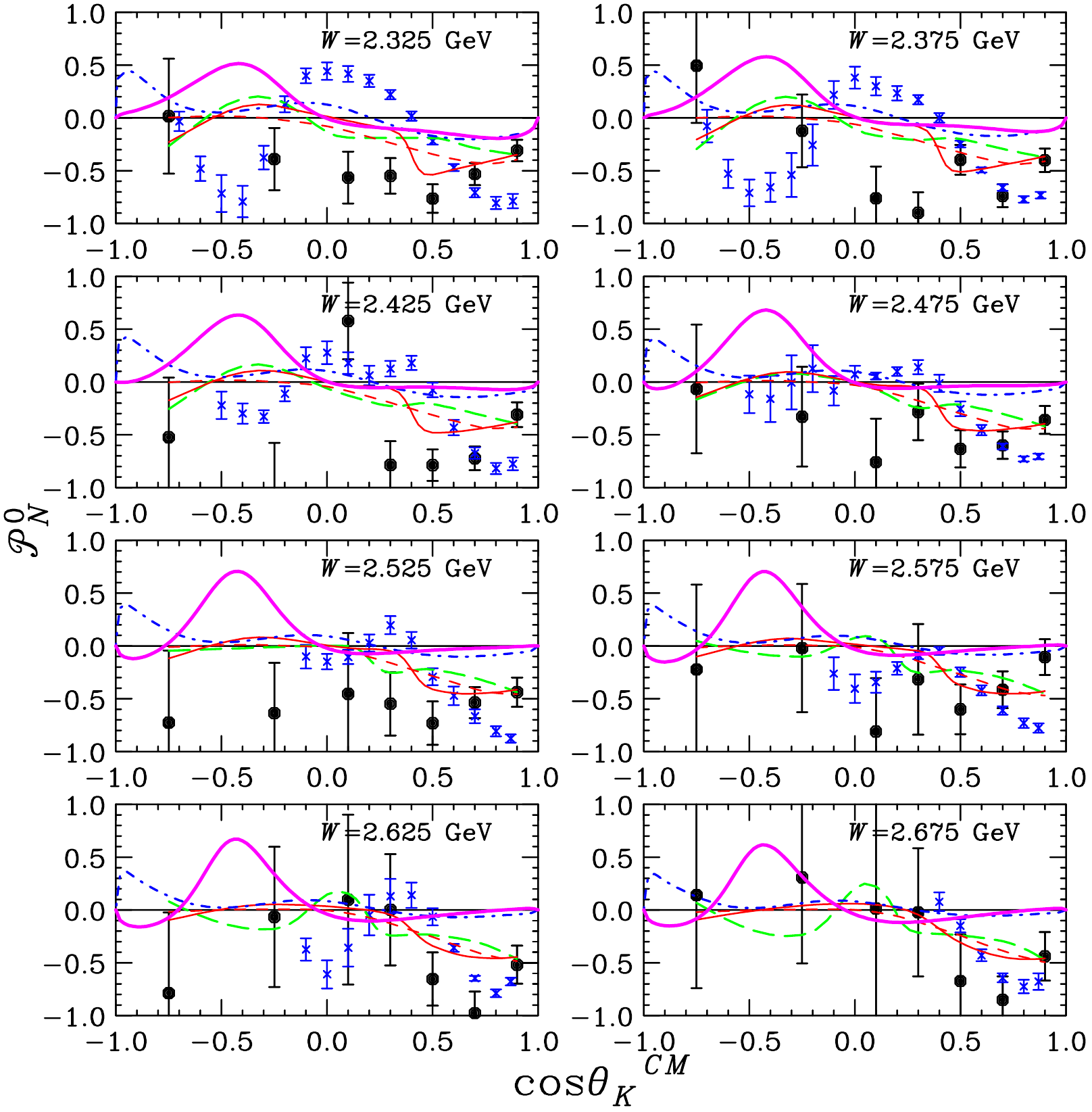}
 \caption[]{(Color online) Same as Fig.~\ref{Pnvscos1} except for $W$ from 2.325 to 2.675~GeV.}
 \label{Pnvscos3}
\end{figure*}

\begin{figure}[tp]
 \centering\includegraphics[height=19cm]{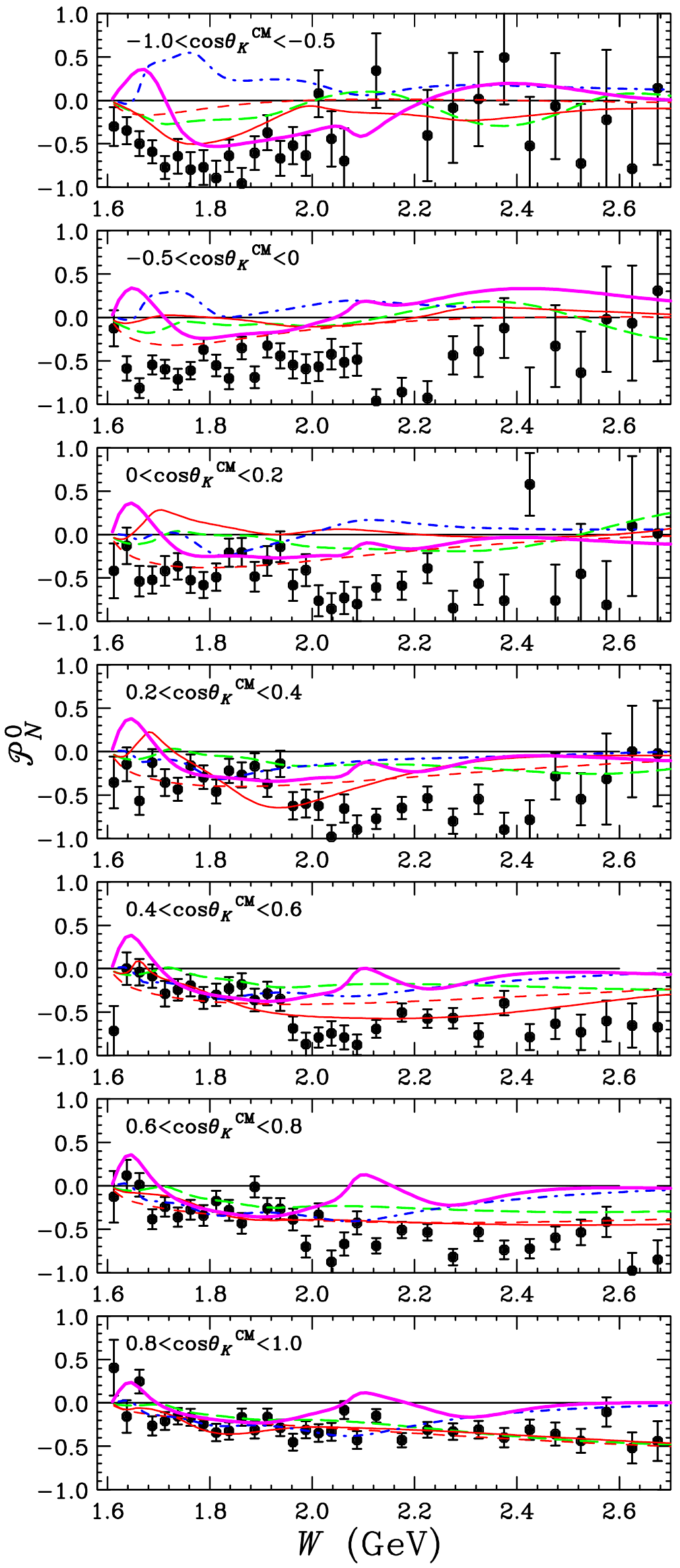}
 \caption[]{(Color online) Induced $\Lambda$ polarization ${\cal P}_N^0$ vs.~$W$ for
   our seven $\cos \theta_K^{CM}$ bins at an average $Q^2$ of 1.90~GeV$^2$. The symbols 
   and curves are the same as Fig.~\ref{Pnvscos1}.}
 \label{PnvsW}
\end{figure}

The $W$ and $\cos\theta_K^{CM}$ dependence of our final data are shown in
Figs.~\ref{Pnvscos1}-\ref{PnvsW} and are available through the CLAS physics
database \cite{CLASDB}. The results are
presented at the geometrical bin centers and not the
event-weighted average of the points. We found that the event-weighted average of
$W$ is identical to the geometrical bin centers to within three
significant figures, while the event-weighted average of
$\cos\theta_K^{CM}$ is generally within $\pm 0.02$ of the geometrical
bin centers.

Figs.~\ref{Pnvscos1}, \ref{Pnvscos2}, and \ref{Pnvscos3} show the induced $\Lambda$
polarization as a function of $\cos\theta_K^{CM}$ along with previous CLAS
photoproduction data~\cite{mccracken}. The average systematic uncertainty on the 
photoproduction points is 0.05. Our electroproduction results are persistently negative,
whereas the photoproduction data are generally positive at backward kaon angles
($\cos\theta_K^{CM}<0$) and negative only for forward kaon angles.
The photoproduction and electroproduction data agree reasonably well for
$\cos\theta_K^{CM}>0.5$ where $t$-channel processes dominate, suggesting a dominant
influence of the transverse component of the virtual photon.  Previous
electroproduction results from CLAS \cite{Raue05,Ambrozewicz:2006zj,Carman:2012}
showed that $\sigma_L$ is small in this kinematic range, even at backward kaon
angles. However, the large difference between the photo- and electroproduction
results seen here suggests that although the longitudinal polarization of the
virtual photon by itself may not play a significant role, even a small
contribution in the interference terms may cause a sizable contribution for this
observable.  Furthermore, although we found a negligible $Q^2$ dependence in our
data, the large differences between the electroproduction and photoproduction
data suggests that somewhere below our lower $Q^2$ limit
(0.8~GeV$^2$) there must be a dramatic change in the electroproduction values of
${\cal P}_N^0$.

%
%

Fig.~\ref{PnvsW} shows the $W$ dependence of ${\cal P}_N^0$ for all $\cos
\theta_K^{CM}$ bins. For the two most forward kaon-angle bins, the variation
with $W$ is smooth,  with no discernible fluctuations other than the monotonic
decrease with increasing $W$. This is consistent with $t$-channel dominance. 
Another feature is that beyond about 2.1~GeV the polarization is essentially constant at a value of -0.5.  In the bins 
from $0<\cos\theta_K^{CM}<0.6$ there is a noticeable fluctuation near 1.9~GeV. 
A resonance structure around 1.9~GeV has been
observed in the photoproduction cross section
\cite{Tran:1998qw,McNabb:2003nf,Bradford:2005pt,mccracken}, as well as in
electroproduction measurements of
$\sigma_T+\epsilon\sigma_L$~\cite{Ambrozewicz:2006zj,Carman:2012} and
$\sigma_{LT'}$~\cite{Nasseripour:2008}. Early work by Bennhold and
Mart~\cite{MB-d13} explained this by postulating contributions from a previously unseen
$J^P=3/2^-$ resonance at 1.96~GeV, although subsequent models and partial wave
analyses~\cite{saghai_aip,Anisovich2012} come to different conclusions. The PDG~\cite{pdg} now lists a three-star, $J^P=3/2^+$ 
resonance that arose from the coupled-channel analysis of Ref.~\cite{Anisovich2012}. Inclusion
of our new induced polarization data in models will be important to better understand the 
contributing $N^*$ states and their coupling parameters.

\subsection{Comparison to Theoretical Models}
\label{pol_theroy}

Our polarization results are compared to three different models.  One is a Regge plus resonance
model (RPR) with three variants, referred to here as RPR-2007~\cite{corthals},
RPR-2011~\cite{tom-vrancx}, and the RPR-2011 model with the resonances turned off~\cite{tom-vrancxPC}, referred to here as
RPR-2011NoRes.  This model treats the non-resonant
background contributions as exchanges of kaonic Regge trajectories in the
$t$-channel, with the $K$ and $K^*$ as the dominant trajectories. To take into
account the $s$-channel contributions, the RPR models include established
$s$-channel nucleon resonances: $N(1650)1/2^-$, $N(1710)1/2^+$, $N(1720)3/2^+$,
as well as the $N(1900)3/2^+$. The older RPR-2007
model was fit to forward-angle ($\cos \theta_K^{CM} > 0$) photoproduction data from
CLAS, LEPS, and GRAAL~\cite{corthals}. The newer RPR-2011 model was fit to the
entire $\cos \theta_K^{CM}$ angular range of all recent $K^+\Lambda$
photoproduction data, including Ref.~\cite{mccracken}. Furthermore it uses a
consistent formalism for the description of spin-5/2 particles as described in
Ref.~\cite{tom-vrancx}. Neither version of the RPR model included any fits to
the existing electroproduction data.

The older RPR-2007 model does a good job of describing the data for $0.8<\cos \theta_K^{CM} < 1$ over
the entire $W$ range (see Fig.~\ref{PnvsW}), but the agreement gets worse as the kaon production angle 
increases. At backward angles it tends to fall somewhere between the
photo- and electroproduction data. The RPR-2011 model does a good job of predicting the data
in our three most forward angle bins but also fails to describe the data as the angle increases. 
It was already noted in Ref.~\cite{Carman:2012} when comparing the RPR-2007 and RPR-2011 models to the
separated structure functions ($\sigma_T+\epsilon \sigma_L$, $\sigma_{TT}$,
$\sigma_{LT}$, and $\sigma_{LT'}$) from this same data set, the RPR-2011
model fares noticeably worse than the RPR-2007 model over all angles for $W< 2.1$~GeV.
To see the effect of resonances on the calculation of this observable, we also include a version of RPR-2011 with 
the resonances turned off, RPR-2011NoRes.   Surprisingly, it seems as though this variant of the model actually 
does a better job of predicting our data over a larger range of kinematics than does RPR-2011. 
Generally speaking, the RPR models best match the data in the forward angle regions where $t$-channel 
processes dominate, but they fail to describe the data at larger angles where nucleon resonances play 
a more significant role.
 
The second model we compare our data to is the Extended Kaon-Maid model~\cite{Mart14}, which was 
originally compared to the low $Q^2$ $K^+\Lambda$ and $K^+\Sigma^0$ data of Ref.~\cite{Achenbach:2011rf}. 
Kaon-Maid \cite{MB-d13} is 
an effective field theory that includes kaon resonances $K^*(892)$ and $K_1(1270)$ in the
$t$-channel, as well as nucleon resonances $N(1650) 1/2^-$, $N(1710) 1/2^+$, $N(1720) 3/2^+$, and 
$N(1895) 3/2^-$, and the extended version used here also includes $N(1675) 5/2^-$, $N(1700) 3/2^-$,
$N(2000) 5/2^+$, and $N(2200) 5/2^-$ (no longer listed in the PDG).  As with the RPR models, the 
Extended Kaon-Maid model generally agrees with the data at forward angles but shows progressively 
worse agreement with the data as the kaon CM angle increases.

Finally, we also include the effective field theory model of Maxwell~\cite{maxwell:12,maxwell:12-2}.
This model was fit to 
all available photo- and electroproduction data (prior to 2012) up to $W=2.3$~GeV.
This model includes contributions from the $t$-channel ($K^*(892)$ and $K_1(1270)$), all three and four
star  $s$-channel resonances up to spin 5/2 from 1440 to 2000~MeV, and several three and four star
$u$-channel resonances with spin up to 5/2. For the range $1.75\leq W\leq 1.95$~GeV, this model fairly 
accurately predicts
the observed $\cos\theta_K^{CM}$ dependence of the data (see Figs.~\ref{Pnvscos1} and \ref{Pnvscos2}). 
However, outside of this narrow range, the model has some fairly obvious deficiencies.  The model predicts a positive
bump near threshold, which grows with angle but is not seen in the data. The model also predicts a fairly 
prominent  bump at around 2.1~GeV (see Fig.~\ref{PnvsW}), which is not seen in the data.  This is likely due to the inclusion of
two resonances ($N(2080)3/2^-$ and $N(2200)5/2^-$) that have recently been removed from the PDG. This model 
has been shown to demonstrate a similarly flat dependence on $Q^2$ as is seen in our data \cite{maxwellPC}, yet it accurately
predicts the results from photoproduction \cite{delaPuente}.  We
are currently working with Maxwell to understand this behavior.

None of the available models does a satisfactory job of describing the
induced $\Lambda$ polarization over the full range of kinematics for our data,
especially at the backward angles where $s$-channel resonances are a larger
part of the overall response. 
Clearly more work on the modeling, and possibly on the fitting/convergence
algorithms, is required to be able to fully understand the contributing $N^* \to
K^+\Lambda$ states and to reconcile the results from the single-channel models
with the currently available coupled-channel models.

\section{Conclusions}
\label{sec:conclusions}

We have presented induced $\Lambda$ polarization results for
$K^+$ electroproduction for a total of 215 ($\cos\theta_K^{CM}$,$W$) bins summed
over $Q^2$ (at an average value of $Q^2$ = 1.90~GeV$^2$),
covering the $W$ range from threshold up to 2.7~GeV and the full kaon
center-of-mass angular range. The induced polarization is uniformly negative, unlike the photoproduction data, which has 
kinematic areas of positive as well as negative polarization.The clear differences with the published CLAS
photoproduction data at the mid and back kaon angles, where $s$-channel
processes become important, emphasize that in studying electroproduction one can learn more about the 
contributing resonant and non-resonant terms. 
Furthermore, given the $Q^2$ independence observed in our data, there must be a 
dramatic change in the production process at lower momentum transfers. It is possible 
that future experiments using CLAS12 at Jefferson Lab may be able to probe this regime.

The data do not clearly indicate any obvious structures in the $W$ dependence that one may interpret as 
indications of strong influences of $s$-channel resonances. However, there are large differences between
the RPR-2011 model with and without resonances at back angles, indicating the importance of including
such terms.  The polarization
data above $W$ = 2~GeV at forward angles are reasonably well described by a non-resonant
Regge mechanism.

At the moment none of the available theoretical models can satisfactorily
explain our results over the full kinematic range of the data. The predictions
of both  RPR theoretical models and the Extended Kaon-Maid model are in fair agreement with the experimental data
at very forward kaon angles, but fare poorly when compared
against the data in the rest of the kinematic phase space.  The Maxwell model works well in the 
range $1.75\leq W\leq 1.95$~GeV over most of the angular range, but is a poor match to the data 
elsewhere. These findings are a
strong indication that these data can be used to provide important constraints
on future model fits, particularly when included within a fully coupled-channel
partial-wave analysis. The sizable differences of the polarization results
between the photo- and electroproduction data in the same $W$ and $\cos
\theta_K^{CM}$ range make clear that for a detailed understanding of the
contributing resonant and non-resonant terms to the $K^+\Lambda$ final state, 
combined fits to both the photo- and
electroproduction data will be essential.  Additionally, measurements of ${\cal
  P}_N^0$ for the $\Sigma^0$ are also important for such fits because the $\Sigma^0$
provides access to additional intermediate states not accessible to the
$\Lambda$. Specifically, it allows access to intermediate $\Delta$ and $\Delta^*$ states with isospin 3/2 in addition to the
isospin 1/2 $N^*$ states that are accessible to the $\Lambda$ final state. Work is 
currently underway to extract ${\cal P}_N^0$ for the $\Sigma^0$ using these same data.
\clearpage

\begin{acknowledgments}
  We are grateful for the efforts of the staff of the Accelerator and Physics
  Divisions at Jefferson Lab that made this experiment possible.  M. Gabrielyan
  was supported in part by a Florida International University Dissertation Year
  Fellowship. 
  This work was supported in part by  the Chilean Comisi\'on Nacional de Investigaci\'on Cient\'ifica y Tecnol\'ogica (CONICYT),
  the Italian Istituto Nazionale di Fisica Nucleare, the French Centre National de la Recherche Scientifique,
  the French Commissariat \`{a} l'Energie Atomique, the U.S. Department of Energy,
  the National Science Foundation, the Scottish Universities Physics Alliance (SUPA),
  the United Kingdom's Science and Technology Facilities Council, and the National Research Foundation of Korea.
  The Southeastern Universities Research Association (SURA) operates the Thomas Jefferson National Accelerator Facility for the United 
  States Department of Energy under contract DE-AC05-84ER40150.
\end{acknowledgments}

\bibliography{IndPol}
\end{document}